\documentclass[lettersize,journal]{IEEEtran}
\usepackage{xcolor}
\usepackage{amsmath,amsfonts}
\usepackage{amssymb}
\usepackage{amsthm}
\newtheorem{proposition}{Proposition}
\usepackage{algorithmic}
\usepackage{algorithm}
\usepackage{array}
\usepackage[font=normalsize,labelfont=sf,textfont=sf]{subcaption}
\usepackage{textcomp}
\usepackage{stfloats}
\usepackage{url}
\usepackage{verbatim}
\usepackage{graphicx}
\usepackage{cite}
\usepackage{subfig}
\usepackage{booktabs}
\theoremstyle{plain}

\theoremstyle{plain}
\newtheorem{rem}{Remark}

\usepackage{amsmath}

\newtheorem{example}{Example}
\begin{document}

\title{DWM-RO: Decentralized World Models with Reasoning Offloading for SWIPT-enabled Satellite-Terrestrial HetNets}

\author{Guangyuan Liu, Yinqiu Liu, Ruichen Zhang, Nan Ma, Jiawen~Kang, Sumei Sun,~\IEEEmembership{Fellow,~IEEE}, \\Abbas Jamalipour,~\IEEEmembership{Fellow,~IEEE}, and Ping Zhang,~\IEEEmembership{Fellow,~IEEE}
\thanks{G.~Liu is with the College of Computing and Data Science, the Energy Research Institute @ NTU, Interdisciplinary Graduate Program, Nanyang Technological University, Singapore (e-mail: liug0022@e.ntu.edu.sg).}
\thanks{Y.~Liu, R.~Zhang are with the College of Computing and Data Science, Nanyang Technological University, Singapore (e-mails: yinqiu001@e.ntu.edu.sg, ruichen.zhang@ntu.edu.sg).}
\thanks{N. Ma and P. Zhang are with the State Key Laboratory of Networking and Switching Technology, Beijing University of Posts and Telecommunications, China (emails: manan@bupt.edu.cn and pzhang@bupt.edu.cn).}
\thanks{J. Kang is with the School of Automation, Guangdong University of Technology, China. (e-mail: kavinkang@gdut.edu.cn).}
\thanks{S. Sun is with the Institute for Infocomm Research, Agency for Science, Technology and Research, Singapore (e-mail: sunsm@i2r.a-star.edu.sg).}
\thanks{A. Jamalipour is with the School of Electrical and Computer Engineering, University of Sydney, Australia, and with the Graduate School of Information Sciences, Tohoku University, Japan (e-mail: a.jamalipour@ieee.org).}
\thanks{G. Liu and Y. Liu contributed equally to the work.}
\vspace{-1cm}
}
\maketitle

\begin{abstract}

Wireless networks are undergoing a paradigm shift toward massive connectivity with energy-efficient operation, driving the integration of satellite-terrestrial architectures with simultaneous wireless information and power transfer (SWIPT). Optimizing transmit beamforming and power splitting in such systems faces formidable challenges, e.g., time-varying channels and multi-tier interference, which create a complex decision landscape. In such a situation, conventional model-free multi-agent reinforcement learning (MARL) suffers from sample inefficiency due to rarely-encountered state transitions and poor coordination as decentralized agents act independently. This paper proposes the Decentralized World Model with Reasoning Offloading (DWM-RO) framework to address these fundamental limitations. An uncertainty-aware offloading gate monitors local interference levels and model reconstruction errors to trigger selective edge coordination, where a lightweight latent decorrelation mechanism refines agents' strategic representations and guides them toward orthogonal actions that minimize resource conflicts. Extensive simulations demonstrate that DWM-RO converges in fewer steps than state-of-the-art baselines while achieving 34.7\% higher spectral efficiency and reducing constraint violations by 40\%. In dense network scenarios with 10 users, DWM-RO maintains violation rates below 20\% while baselines exceed 70\%, validating superior robustness.
\end{abstract}


\vspace{-0.2cm}
\section{Introduction}
Beyond 6G wireless communications are expected to revolutionize global connectivity by accommodating an unprecedented number of devices~\cite{chu2025revolutionizing,11195786,9390169}. 
However, achieving high spectral efficiency while ensuring energy sustainability for battery-constrained terminals remains a key challenge~\cite{zhang2021joint,9502719}.
Integrated satellite-terrestrial heterogeneous networks (HetNets) have emerged as a promising architectural solution to address these dual objectives~\cite{pan2017performance}. By using high-power satellites to provide wide-area coverage and deploying low-power terrestrial femto base stations (FBSs) to enhance local capacity, such networks offer improved scalability and spatial reuse.
Satellites, characterized by strong long-range broadcasting capabilities and global coverage, are particularly suitable for backhaul provisioning or service continuity in remote and disaster-prone areas.
To further improve spectral utilization, satellite and terrestrial tiers are typically configured to operate in a co-frequency mode~\cite{zhang2021joint,pan2022space}. 
Despite improving spectrum reuse, this configuration also creates a complex multi-tier interference environment. 
In particular, cross-tier interference, where satellite downlink transmissions affect terrestrial access links, becomes a significant bottleneck that demands advanced coordination and mitigation~\cite{10032267}.

In parallel with spectrum reuse, simultaneous wireless information and power transfer (SWIPT) has emerged as an effective technique to improve energy sustainability in wireless networks~\cite{pan2017performance, pan2016secrecy, chu2025revolutionizing}.
SWIPT enables wireless devices to simultaneously decode information and harvest energy from ambient radio frequency (RF) signals~\cite{pan2022space}.
The most typical SWIPT architecture is power splitting (PS), which divides the received signal for concurrent information decoding and energy harvesting (EH)~\cite{pan2017performance,zhang2021joint,pan2022space}. 
When integrated into satellite-terrestrial HetNets, SWIPT offers synergistic advantages. In particular, the reduced transmission distances within femtocells significantly enhance RF EH efficiency~\cite{clerckx2018fundamentals, 10354077,pan2022space}.
However, combining SWIPT with satellite-terrestrial HetNets introduces significant system-level complexity. In particular, the joint optimization of transmit beamforming and power splitting ratios becomes a critical challenge due to the intertwined interference and EH dynamics.
This problem is further compounded in highly dynamic wireless environments, where time-varying channels, mobility, and stochastic interference make the optimization landscape non-stationary.
Traditional convex optimization techniques, which rely on accurate and often centralized channel state information (CSI) and assume relatively static conditions, are inadequate for such scenarios~\cite{ng2014robust, liu2024survey}. {
This challenge motivates a shift toward the convergence of communication and AI (ComAI) paradigm, which calls for natively embedding AI to optimize overall system performance and create an intelligent, adaptive network ecosystem, rather than optimizing isolated modules~\cite{Zhang2025ComAI,ni2025llm}.
As a result, there is a growing need for scalable, adaptive optimization frameworks that can operate effectively under limited information and real-time variations in SWIPT-enabled satellite-terrestrial networks.}

These limitations have spurred growing interest in learning-based optimization approaches, particularly multi-agent reinforcement learning (MARL)~\cite{feriani2021single, 8792117}, which offers a natural paradigm for distributed resource allocation in complex and dynamic wireless systems.
Recent studies have demonstrated the feasibility of MARL in various HetNet scenarios. For example, the authors in \cite{8792117}, \cite{10322786}, and \cite{10938906} applied multi-agent deep Q-learning, multi-agent proximal policy optimization (MAPPO), and multi-agent deep deterministic policy gradient (MADDPG) for dynamic power allocation, resource management, and network slicing in HetNets, showing notable performance gains. 
Nonetheless, existing MARL frameworks face two fundamental challenges when applied to SWIPT-enabled satellite-terrestrial HetNets.
\begin{itemize}
    \item \textbf{Challenge I: Prohibitive Sample Inefficiency.} Conventional MARL algorithms, such as MAPPO~\cite{kang2023cooperative}, often suffer from sample inefficiency~\cite{MABL, liu2024efficient}, requiring tremendous interactions with the environment to learn effective policies. This issue becomes more critical in SWIPT-enabled satellite-terrestrial HetNets due to the inherent complexity of the environment, including cross-tier interference, time-varying fading, and nonlinear EH dynamics. Such rich and diverse state transitions are rarely encountered during limited training, preventing model-free MARL from sufficiently exploring the vast state space. Consequently, the learning process is significantly delayed, and the resulting policies often fail to generalize to underrepresented conditions, such as unexpected interference surges.
    
    \item \textbf{Challenge II: Lack of Coordination.} In standard decentralized MARL settings, each agent operates based solely on its own local observations, without knowledge of other agents' actions or observations~\cite{MABL, nomura2025decentralized}. This information asymmetry prevents coordination, often leading to resource conflicts. For instance, in SWIPT-enabled satellite-terrestrial HetNets, when multiple agents simultaneously experience local interference, they may each independently increase their transmission power. However, without coordination, such responses collectively escalate interference and create a feedback loop that degrades overall network performance. 
\end{itemize}

In this paper, we present the \underline{D}ecentralized \underline{W}orld \underline{M}odel with \underline{R}easoning \underline{O}ffloading (DWM-RO) framework. \textit{To the best of our knowledge, this is the first work to apply decentralized world models to optimize satellite-terrestrial HetNets.} 
Unlike conventional model-free MARL that requires extensive interaction with the environment, world models enable agents to learn compact and predictive latent representations of environment dynamics~\cite{hafner2019dream,hafnerdiverse,hafner2025nature}, allowing them to perform imagination-based planning through internally simulated trajectories. 
{
To further enhance inter-agent coordination, we propose a reasoning offloading mechanism in which agents selectively offload their latent representations to an edge server for edge-assisted latent refinement. In this context, ``reasoning'' refers to the decision-relevant latent belief learned by the RSSM, rather than an explicit symbolic or language-based reasoning process. The edge server performs latent-space decorrelation to resolve potential resource conflicts and refines agent strategies accordingly. In this way, DWM-RO establishes an efficient and robust paradigm for distributed optimization in highly dynamic and interference-limited wireless environments. The major contributions of this paper can be summarized as follows.}
\begin{itemize}
    \item \textbf{SWIPT in Satellite-Terrestrial HetNets}: We develop the system model for SWIPT-enabled satellite-terrestrial HetNets that incorporates realistic physical-layer characteristics, including non-linear EH, time-varying channels, and multi-tier interference. We formulate the joint optimization problem of transmit beamforming and power splitting ratios. 
    \item \textbf{Decentralized World Model}: We present DWM-RO that synergistically integrates decentralized world models with adaptive offloading. First, each MARL agent independently learns a predictive model that compresses high-dimensional observations into compact latent representations, capturing both deterministic temporal dependencies and stochastic uncertainties in the wireless environment. The world model architecture enables agents to perform imagination-based policy optimization through simulated rollouts entirely within the latent space. Such a model-based MARL fundamentally addresses \textbf{Challenge I} by dramatically improving sample efficiency.
    \item \textbf{Uncertainty-Aware Offloading}: Moreover, we propose an adaptive coordination mechanism comprising a learnable offloading gate coupled with a lightweight latent decorrelation strategy. The gate monitors local uncertainty indicators (e.g., interference levels and world model reconstruction errors) to make real-time decisions on whether to trigger edge assistance. When offloading is activated, the edge server performs efficient latent refinement by removing the shared strategic component from agents' latent states and thus producing decorrelated representations that guide agents toward orthogonal action selections. This mechanism directly addresses \textbf{Challenge II} by resolving resource conflicts.
\end{itemize}

The remainder of this paper is organized as follows. Section II reviews the related work. Section III presents the system model and problem formulation. Section IV details the DWM-RO framework. Section V presents the simulation results. Finally, Section VI concludes the paper.

\section{Related Work}
\subsection{SWIPT-Enabled Satellite-Terrestrial HetNets}
SWIPT-enabled HetNets have attracted great research attention because of their potential to jointly enhance spectral efficiency and energy sustainability.
For instance, Zhang~\textit{et al.}~\cite{8891923} proposed an energy-efficient resource management scheme in NOMA-based HetNets with SWIPT, where the joint power allocation and subchannel matching problem was decoupled and solved iteratively.
In~\cite{8840901}, the authors formulated an energy efficiency maximization problem that considered both bounded CSI uncertainty and a nonlinear EH model. 
A minimax probability machine method was employed to convert the probabilistic outage constraints into deterministic convex forms.
In the context of reinforcement learning, Zhang~\textit{et al.}~\cite{zhang2021joint} addressed the joint optimization of coordinated beamforming and power splitting by proposing a multi-agent deep Q-network (DDQN) approach.
\begin{figure*}[t!]
    \centering
    \includegraphics[width=0.85\textwidth]{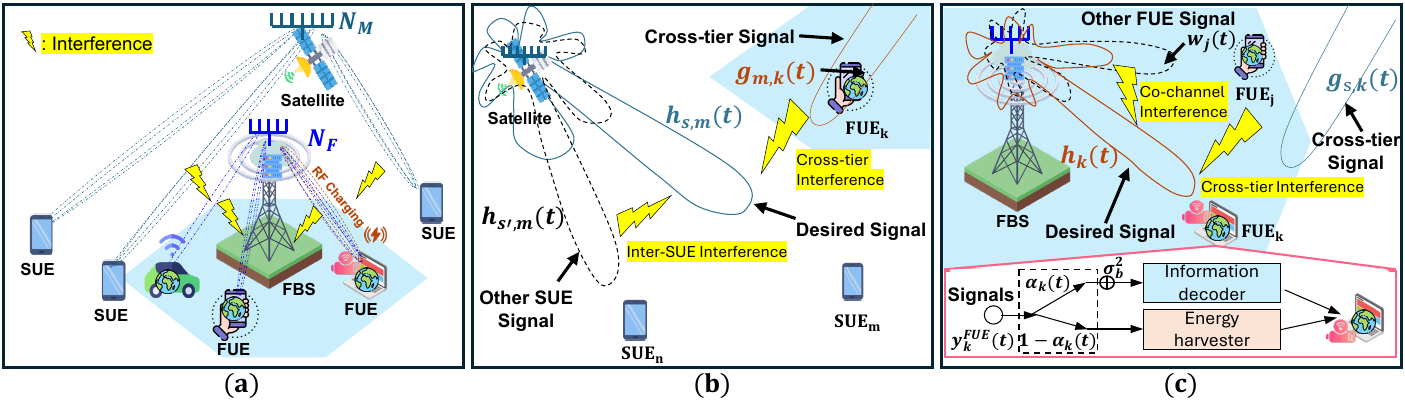}
\caption{\footnotesize
    An illustration of the considered SWIPT-enabled satellite-terrestrial HetNet model. 
    \textbf{(a)} The overall network topology.
    \textbf{(b)} The communication model for each SUE. It receives a desired signal from the satellite, along with inter-SUE interference from other satellite beams and cross-tier interference from the FBS~\cite{sharma2017performance,10032267}.
    \textbf{(c)} The communication and receiver model for one FUE. The FUE receives a desired signal and co-channel interference from the FBS, in addition to cross-tier interference from the satellite. The inset details its PS circuit, where the signal is split for ID (with processing noise $\sigma_{b}^{2}$) and EH.
    }
    \label{fig:system_model}
    \vspace{-0.4cm}
\end{figure*}

However, these studies primarily focused on terrestrial HetNets and did not consider the unique challenges introduced by integrating satellites.
Satellite-terrestrial HetNets present fundamentally different characteristics, such as complex multi-tier interferences between satellite and terrestrial networks operating in co-frequency mode~\cite{10159432,wang2024safeguarding}.
Recent work has explored SWIPT in satellite-terrestrial systems. For example, Wang~\textit{et al.} \cite{10159432} investigated robust secure beamforming and power splitting for millimeter-wave cognitive satellite-terrestrial networks with SWIPT, considering imperfect CSI. Zhang \textit{et al.} \cite{10354077} analyzed secrecy outage performance for UAV-assisted satellite-terrestrial SWIPT systems with NOMA under randomly distributed eavesdroppers.
Despite these advances, existing proposals predominantly rely on model-based 
optimization techniques such as robust beamforming~\cite{10159432} and stochastic geometry-based analysis~\cite{10354077}. Such approaches cannot adapt to the highly dynamic nature of satellite-terrestrial HetNets with time-varying channels, mobility, and evolving interference patterns. These limitations motivate the use of MARL approaches, especially world models.

\subsection{World Models in Communication Optimization}
{World models have emerged as a powerful paradigm in ComAI by enabling agents to learn compact latent representations of network dynamics and perform sample-efficient planning through imagination~\cite{ha2018world , Zhang2025ComAI}.}
For example, Wang~\textit{et al.}~\cite{worldmodel1} proposed a world-model-based reinforcement learning framework for vehicular networks, where the model predicts the long-term evolution of the age of information (AoI) under dynamic channels to support trajectory planning without extensive real-world interactions.
Chai~\textit{et al.}~\cite{worldmodel3} introduced MobiWorld, a generative world model that integrates heterogeneous and multimodal data sources to support high-fidelity simulation environments for communication-aware network planning.
Zhao~\textit{et al.}~\cite{worldmodel2} applied world models to low-altitude economy networks, demonstrating their effectiveness in managing partial observability and temporal dependencies in aerial communication systems through latent imagination and predictive modeling. While these studies illustrate the potential of world models in various networking contexts, they primarily focus on single-agent learning or homogeneous network environments. In contrast, multi-agent coordination in SWIPT-enabled satellite-terrestrial HetNets involves additional challenges, such as decentralized learning under partial observability, complex inter-agent interference, and communication-constrained collaboration. This motivates the need to explore how distributed world models can be adapted to multi-agent settings with communication-aware coordination mechanisms.

{
In comparison, prior learning-based solutions differ mainly in three aspects. First, in terms of sample efficiency, model-free MARL methods~\cite{8792117,10322786,10938906} typically require extensive environment interaction, whereas world-model-based methods~\cite{worldmodel1,worldmodel2,worldmodel3} are more sample-efficient due to latent imagination. Second, in terms of coordination, conventional decentralized MARL methods provide little or no explicit coordination during execution, while existing world-model-based communication studies mainly consider single-agent or homogeneous settings and therefore do not address decentralized inter-agent conflict resolution. Third, in terms of computational characteristics, model-free MARL is relatively simple per step but often incurs long training time, whereas world models introduce additional predictive modules in exchange for substantially reduced interaction cost. The proposed DWM-RO differs from existing approaches along all three dimensions. Specifically, it improves sample efficiency through decentralized world models, strengthens coordination through adaptive edge-assisted latent refinement, and controls online computational and communication overhead by activating coordination only when necessary.
}

\section{System Model and Problem Formulation}\label{sec:system_model}
In this section, we present the system model, modeling wireless communications, EH, and communication channels. Moreover, we formulate the joint optimization problem of beamforming and PS. 

\subsection{System Overview}
As illustrated in Fig.~\ref{fig:system_model}(a), we consider a downlink satellite-terrestrial HetNet with SWIPT. The network comprises a satellite communication system equipped with $N_M$ antennas serving $M$ satellite user equipment (SUE) and a terrestrial femtocell network. The femtocell network consists of one femto base station (FBS) equipped with $N_F$ antennas and serves $K$ femto user equipment (FUE). The satellite system and the femtocell network share the same frequency spectrum to maximize spectral efficiency. Each FUE is equipped with a PS receiver \cite{zhang2021joint,pan2017performance}, enabling simultaneous information decoding (ID) and EH from the received RF signals. At each time slot $t$, FUE $k$ associated with FBS receives downlink signals and must determine a power splitting ratio $\alpha_{k}(t) \in [0,1]$, allocating $\alpha_{k}(t)$ portion of the received power for ID and $(1-\alpha_{k}(t))$ portion for EH. To fully utilize the limited frequency band resources, FBSs use the same licensed spectrum of the satellite without adverse impact on SUEs.

Without loss of generality, we assume all SUEs and FUEs are single-antenna devices. With co-channel spectrum sharing between the satellite and terrestrial tiers and space-division multiple access (SDMA)-based spatial multiplexing (e.g., satellite multibeam downlinks and multiuser precoding at femtocells), multi-tier interference arises. Specifically, FUEs suffer intra-cell (co-channel) interference from other FUEs within the same femtocell and cross-tier interference from satellite downlink beams intended for SUEs, while SUEs are affected by terrestrial femtocell transmissions~\cite{zhang2021joint,an2016secure,zhang2024spectrumnet}.
Similarly, SUEs experience interference from femtocell transmissions. This spectrum reuse creates significant interference management challenges.
 
First, we model the received signals at both SUEs and FUEs, taking into account complex interference~\cite{sharma2017performance,zhang2021joint}.

\subsubsection{Received Signal at SUE}

Consider SUE $m \in \mathcal{M} = \{1, 2, \ldots, M\}$ in time slot $t$. The received signal at SUE $m$ can be given by
\begin{align}
y_m^{\text{SUE}}(t) = &\quad \underbrace{\mathbf{h}_{s,m}^H(t)\mathbf{v}_m(t)x_m(t)}_{\text{desired signal}} \nonumber \\
&+ \underbrace{\sum_{m' \neq m} \mathbf{h}_{s,m'}^H(t)\mathbf{v}_{m'}(t)x_{m'}(t)}_{I_m^{\text{sat}}(t): \;\text{inter-SUE interference} } \nonumber \\
&+ \underbrace{\sum_{k \in \mathcal{K}} \mathbf{g}_{m,k}^H(t)\mathbf{w}_k(t)s_k(t)}_{\!\!\!\!\!\!\!\!\!\!\!\!\!\!I_m^{\text{ter}}(t): \;\text{cross-tier interference from femtocell}\!\!\!\!\!\!\!\!\!\!\!\!\!\!}+ z_m^{\text{SUE}}(t),\label{eq:sue_signal}
\end{align}
where $\textbf{h}_{s,m}(t)$ denotes the channel vector from satellite $s$ to SUE $m$, capturing both path loss and small-scale fading effects \cite{zhang2024generative,zhou2023aerospace}. The satellite employs beamforming with coefficient $\textbf{v}_m(t)$ to serve SUE $m$, and $x_m(t)$ represents the transmitted symbol intended for SUE $m$ with unit power, i.e., $\mathbb{E}[|x_m(t)|^2] = 1$. Inter-SUE interference arises from simultaneous transmissions to other SUEs, where $\textbf{h}_{s',m}(t)$ denotes the channel gain from satellite beam $s'$ (serving SUE $m'$) to SUE $m$~\cite{zhang2024generative}. The cross-tier interference originates from the terrestrial femtocell, where $\mathbf{g}_{m,k}(t) \in \mathbb{C}^{N_F \times 1}$ represents the channel vector from the FBS to SUE $m$, and $\mathbf{w}_k(t) \in \mathbb{C}^{N_F \times 1}$ is the beamforming vector employed by the FBS to serve FUE $k$ with transmitted symbol $s_k(t)$ and $z_m^{\text{SUE}}(t) \sim \mathcal{CN}(0, \sigma_a^2)$ denotes the additive white Gaussian noise (AWGN) at the SUE $m$. The signal model at the SUE is illustrated in Fig.~\ref{fig:system_model}(b).

\subsubsection{Received Signal at FUE}

Similarly, for FUE $k \in \mathcal{K} = \{1, 2, \ldots, K\}$ equipped with a PS receiver, the received signal in the time slot $t$ is given by
\begin{align}
y_k^{\text{FUE}}(t) = &\quad \underbrace{\mathbf{h}_k^H(t)\mathbf{w}_k(t)s_k(t)}_{\text{desired signal}} \nonumber \\
&+ \underbrace{\sum_{j \neq k, j \in \mathcal{K}} \mathbf{h}_k^H(t)\mathbf{w}_j(t)s_j(t)}_{I_k^{\text{co}}(t): \;\text{co-channel interference}} \nonumber \\
&+ \underbrace{\sum_{m \in \mathcal{M}} \mathbf{g}_{s,k}^H(t)\mathbf{v}_m(t)x_m(t)}_{\!\!\!\!\!\!\!\!\!I_k^{\text{sat}}(t):\;\text{cross-tier interference from satellite}\!\!\!\!\!\!\!\!\!} + z_k^{\text{FUE}}(t),\label{eq:fue_signal}
\end{align}
where $\mathbf{h}_k(t) \in \mathbb{C}^{N_F \times 1}$ denotes the channel vector from FBS to FUE $k$. The FBS employs the beamforming vector $\mathbf{w}_k(t)$ to transmit symbol $s_k(t)$ to FUE $k$, subject to the power constraint $\|\mathbf{w}_k(t)\|^2 \leq P_{\max}$. The co-channel interference arises from simultaneous transmissions to other FUEs served by the same FBS~\cite{clerckx2018fundamentals}. Since our system consists of a single femtocell, there is no inter-cell interference, distinguishing our scenario from multi-cell deployments. Cross-tier interference originates from satellite downlinks~\cite{li2018robust}, where $\textbf{g}_{s,k}(t)$ represents the channel vector from the satellite to the FUE $k$. Since the received antenna noise power is much smaller than the AWGN of the baseband processing circuit power in practice, the received antenna noise power can be ignored~\cite{lohani2016downlink,lu2018coordinated,zhang2021joint}. Thus, $z_k^{\text{FUE}}(t) \sim \mathcal{CN}(0, \sigma_b^2)$ models the noise introduced by the baseband processing circuit at the FUE with variance $\sigma_b^2$. The architecture of the FUE power splitting receiver is illustrated in Fig.~\ref{fig:system_model}(c).

\subsubsection{SINR Analysis}

Based on the received signal models, we can characterize the signal-to-interference-plus-noise ratio (SINR) for both user types. For SUE $m$, the SINR $\Gamma_m^{\text{SUE}}(t)$ is given by \cite{zhang2021joint}
\begin{equation}
\Gamma_m^{\text{SUE}}(t) = \frac{|\mathbf{h}_{s,m}^H(t)\mathbf{v}_m(t)|^2}{I_m^{\text{sat}}(t) + I_m^{\text{ter}}(t) + \sigma_a^2},
\end{equation}
where $I_m^{\text{sat}}(t)$ and $I_m^{\text{ter}}(t)$ denote inter-SUE interference and cross-tier interference from femtocell, respectively, as defined in \eqref{eq:sue_signal}.
For FUE $k$ with PS ratio $\alpha_k(t) \in [0,1]$, the effective SINR for ID can be defined as~\cite{zhang2013mimo}
\begin{equation}
\Gamma_k^{\text{FUE}}(t) = \frac{\alpha_k(t)|\mathbf{h}_k^H(t)\mathbf{w}_k(t)|^2}{\alpha_k(t)(I_k^{\text{co}}(t) + I_k^{\text{sat}}(t)) + \sigma_b^2},
\end{equation}
where $I_k^{\text{co}}(t)$ and $I_k^{\text{sat}}$ denote co-channel interference and the cross-tier interference from satellite are defined in \eqref{eq:fue_signal}, respectively.

Note that the PS operation affects both the desired signal power and the interference power proportionally, while noise power remains constant~\cite{zhang2013mimo,zhang2021joint}.
Consequently, the achievable information rates for SUE $m$ and FUE $k$ are respectively calculated as
\begin{equation}
R_m^{\text{SUE}}(t) = \log_2(1 + \Gamma_m^{\text{SUE}}(t)),
\end{equation}
and
\begin{equation}
R_k^{\text{FUE}}(t) = \log_2(1 + \Gamma_k^{\text{FUE}}(t)).\label{eq:fue_rate}
\end{equation}

\subsection{Energy Harvesting Model}

The received signal at each FUE is split into two parts: $\sqrt{\alpha_k(t)}$ portion is used for ID, and $\sqrt{1-\alpha_k(t)}$ portion is used for EH. Therefore, the power allocated to EH at FUE $k$ can be given by 
\begin{equation}
P_k^{\text{EH}}(t) = (1 - \alpha_k(t)) ( |\mathbf{h}_k^H(t)\mathbf{w}_k(t)|^2 + I_k^{\text{co}}(t) + I_k^{\text{sat}}(t) ).
\end{equation}

{
For the traditional linear EH model, the harvested power is $\delta_k P_k^{\text{EH}}(t)$, where $\delta_k \in (0,1]$ refers to the RF-to-DC conversion efficiency \cite{clerckx2018fundamentals}. However, the linear model cannot capture the nonlinear characteristics of practical EH circuits. To address this mismatch, we adopt the logistic-function-based nonlinear EH model~\cite{zhang2021joint}, where the harvested power at FUE $k$ is defined as
\begin{equation}
  E_k(t)=\frac{\displaystyle\frac{E_{\max}}{1+\mathrm{e}^{-\mu\!(P_k^{\text{EH}}(t)-\nu)}}-\displaystyle\frac{E_{\max}}{1+\mathrm{e}^{\mu\nu}}}
               {1-\displaystyle\frac{1}{1+\mathrm{e}^{\mu\nu}}}, 
  \label{eq:EH_nl}
\end{equation}
where $E_{\max}$ denotes the maximum harvestable power when the EH circuit reaches saturation, $\mu$ and $\nu$ are circuit-specific parameters representing the sensitivity and turn-on threshold of the EH circuit \cite{zhang2021joint}, respectively. For simplicity, the numerical results in this paper use common EH parameters across FUEs. The model in~\eqref{eq:EH_nl} can nevertheless be extended directly to heterogeneous devices by introducing user-specific parameters $\{E_{\max,k}, \mu_k, \nu_k\}$, which can also be included in the local observation as static context for policy adaptation.
}

\subsection{Time-Varying Channel Model}
To capture the dynamic nature of the wireless environment, we model each channel in the network as a product of a large-scale path loss component and a small-scale fading component. The specific models differ for satellite-to-ground and terrestrial links to reflect their distinct propagation characteristics.

\subsubsection{Satellite-to-Ground Links}
The links originating from the satellite, including the channel to SUE $m$, $\mathbf{h}_{s,m}(t)$, and the cross-tier channel to FUE $k$, $\mathbf{g}_{s,k}(t)$, as first introduced in \eqref{eq:sue_signal} and \eqref{eq:fue_signal}, are respectively given by
\begin{align}
    \mathbf{h}_{s,m}(t) &= \sqrt{\beta_{s,m}} \cdot \tilde{\mathbf{g}}_{s,m}(t), \\
    \mathbf{g}_{s,k}(t) &= \sqrt{\beta_{s,k}} \cdot \tilde{\mathbf{g}}'_{s,k}(t),
\end{align}
where $\beta_{s,m}$ and $\beta_{s,k}$ represent the large-scale path loss for the satellite-to-SUE and satellite-to-FUE links, respectively. For satellite links, this path loss is determined by the free-space path loss formula, incorporating specific user antenna gains~\cite{10949621,yue2023low}, i.e.,
\begin{align}
    \beta_{s,m}(d_{s,m}) &= G_s G_m \left(\frac{c}{4\pi f_c d_{s,m}}\right)^2, \\
    \beta_{s,k}(d_{s,k}) &= G_s G_k \left(\frac{c}{4\pi f_c d_{s,k}}\right)^2,
\end{align}
where $G_s$ is the satellite antenna gain, while $G_m$ and $G_k$ are the antenna gains for SUE $m$ and FUE $k$. Additionally, $c$ is the speed of light, $f_c$ is the carrier frequency, and $d$ represents the link distance. The terms $\tilde{\mathbf{g}}_{s,m}(t)$ and $\tilde{\mathbf{g}}'_{s,k}(t)$ are the small-scale fading vectors, modeled with a Rician distribution to account for the strong line-of-sight (LoS) path typical in satellite communications~\cite{10949621}.

\subsubsection{Terrestrial Links}
Similarly, the links originating from the $N_F$-antenna FBS its associated FUE $k$ are modeled as $\mathbf{h}_k(t) \in \mathbb{C}^{N_F \times 1}$, and the cross-tier interference channel from the FBS to SUE $m$, denoted by $\mathbf{g}_{m,k}(t) \in \mathbb{C}^{N_F \times 1}$. These channels are respectively defined as
\begin{align}
    \mathbf{h}_k(t) &= \sqrt{\beta_k} \cdot \tilde{\mathbf{g}}_k(t), \\
    \mathbf{g}_{m,k}(t) &= \sqrt{\beta_{m,k}} \cdot \tilde{\mathbf{g}}'_{m,k}(t),
\end{align}
where $\beta_k$ and $\beta_{m,k}$ are the large-scale path loss components based on a standard terrestrial channel model. $\tilde{\mathbf{g}}_k(t)$ and $\tilde{\mathbf{g}}'_{m,k}(t)$ are the small-scale fading vectors, modeled with a Rayleigh distribution for terrestrial links that may lack a dominant LoS path~\cite{zhang2021joint}.

\subsubsection{Temporal Dynamics}
The time-varying nature of all channels is captured in the evolution of their respective small-scale fading vectors (e.g., $\tilde{\mathbf{g}}_{s,m}(t)$, $\tilde{\mathbf{g}}_k(t)$). Following the Jakes fading model~\cite{xiao2002second}, these components evolve according to a first-order complex Gauss-Markov process, i.e.,
\begin{equation} \label{eq:jakes_fading}
    \mathbf{g}(t) = \rho \mathbf{g}(t-1) + \sqrt{1-\rho^2} \mathbf{e}(t),
\end{equation}
where $\mathbf{g}(t)$ represents any of the small-scale fading vectors, $\mathbf{g}(0) \sim \mathcal{CN}(0,\mathbf{I})$ and $\mathbf{e}(t) \sim \mathcal{CN}(0,\mathbf{I})$ are the initial fading state and the innovation process, with dimensions appropriate for the link ($N_M \times 1$ or $N_F \times 1$). The temporal correlation coefficient $\rho$ is determined by
\begin{equation}\label{eq:jakes_rho}
    \rho = J_0(2\pi f_d T_s),
\end{equation}
where $J_0(\cdot)$ is the zeroth-order Bessel function of the first kind, $f_d$ is the maximum Doppler frequency, and $T_s$ is the time slot duration.

The time-varying and link-dependent nature of these channels poses significant challenges for resource allocation. Channel conditions can change rapidly, making it difficult to obtain accurate global CSI at a centralized controller. Moreover, the temporal correlation implies that current decisions affect future system performance, necessitating predictive capabilities for effective resource management.


\subsection{Problem Formulation}
Our objective is to maximize the spectrum efficiency of the femtocell network while guaranteeing the quality of service (QoS) requirements for both satellite and terrestrial users.
Specifically, at each time slot $t$, we jointly optimize the transmit beamforming vectors $\{\mathbf{w}_k(t)\}_{k \in \mathcal{K}}$ at the FBS and the power splitting ratios $\{\alpha_k(t)\}_{k \in \mathcal{K}}$ at the FUEs. The optimization problem is formulated as
\begin{subequations}\label{eq:main_prob}
\begin{align}
  \mathcal{P}_1: \quad \underset{\{\mathbf{w}_k(t),\alpha_k(t)\}}{\max}\;&
        \sum_{k\in\mathcal{K}} R_k^{\text{FUE}}(t)              \label{eq:obj}\\[2pt]
  \text{s.t.}\quad
       & R_m^{\text{SUE}}(t) \ge \xi_{\text{SUE}},  && \forall m\in\mathcal{M}, \label{eq:sue_qos}\\
       & R_k^{\text{FUE}}(t) \ge \xi_{\text{FUE}},  && \forall k\in\mathcal{K}, \label{eq:fue_qos}\\
       & E_k(t)            \ge \Phi_{\text{FUE}},   && \forall k\in\mathcal{K}, \label{eq:eh_constraint}\\
       & \|\mathbf{w}_k(t)\|^2 \le P_{\max},        && \forall k\in\mathcal{K}, \label{eq:power}\\
       & 0 \le \alpha_k(t) \le 1,                   && \forall k\in\mathcal{K}, \label{eq:ps_range}
\end{align}
\end{subequations}
where constraint~\eqref{eq:sue_qos} ensures that the achievable rate of each SUE meets its minimum QoS requirement $\xi_{\text{SUE}}$, protecting satellite communication from excessive cross-tier interference. The constraint~\eqref{eq:fue_qos} guarantees the minimum rate requirement $\xi_{\text{FUE}}$ for each FUE. The constraint~\eqref{eq:eh_constraint} ensures that each FUE harvests sufficient energy $\Phi_{\text{FUE}}$ to sustain its operation, where $E_k(t)$ follows the nonlinear EH model in~\eqref{eq:EH_nl}. The constraint~\eqref{eq:power} limits the transmit power for each beamforming vector to $P_{\max}$, and constraint~\eqref{eq:ps_range} ensures that the PS ratios are within the valid range.

\begin{rem}
$\mathcal{P}_1$ is inherently non-convex due to the coupled optimization variables in the SINR expressions and the nonlinear EH constraints~\cite{9502719}.
The non-convexity of $\mathcal{P}_1$, combined with time-varying channel dynamics and multi-tier interference, precludes conventional optimization approaches that assume static global CSI. Moreover, the scale of modern femtocell deployments demands distributed solutions that can adapt to local conditions while maintaining system-wide coordination.
\end{rem}

\begin{figure*}
    \centering
    \includegraphics[width=0.9\linewidth]{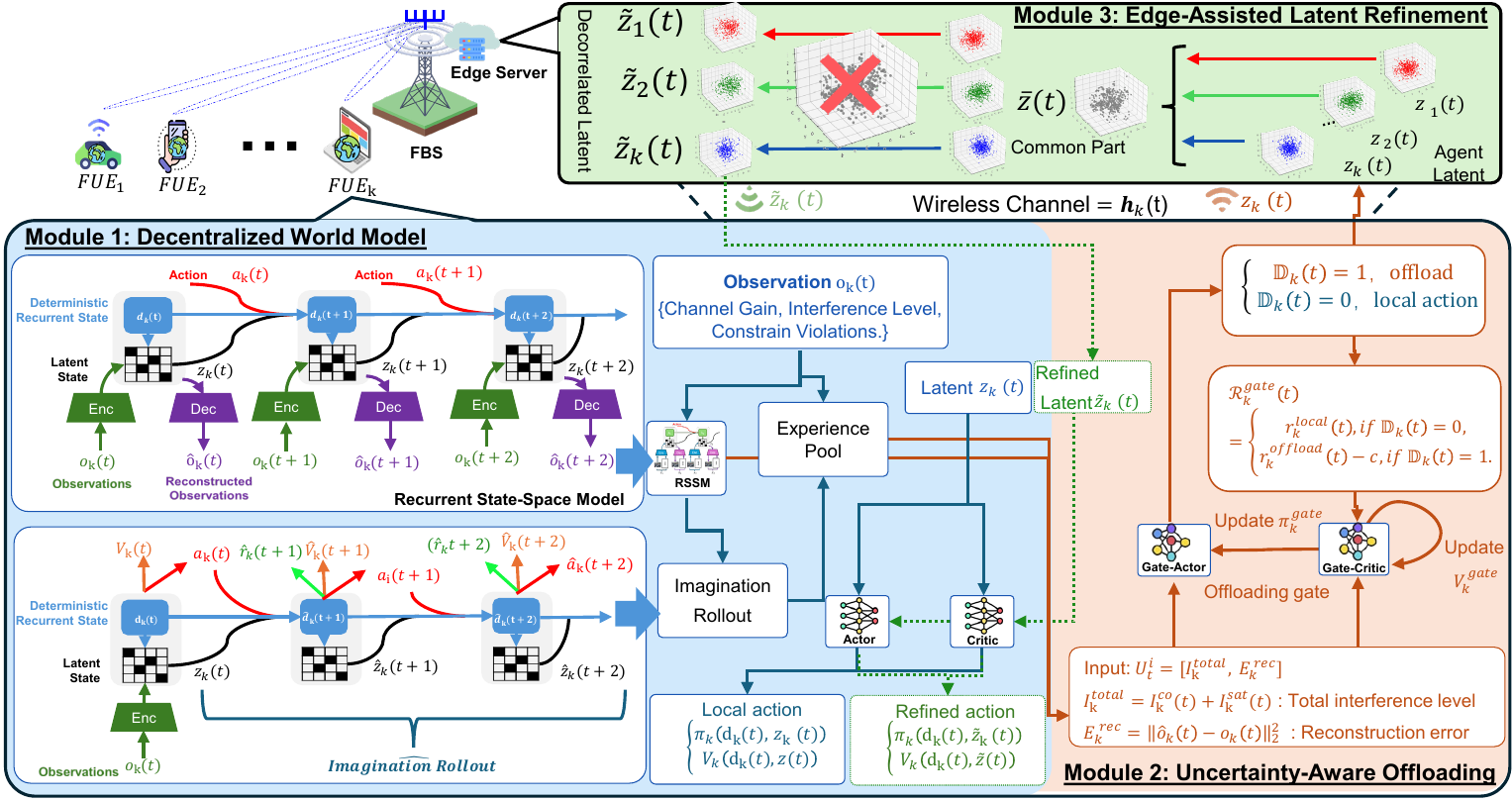}
\footnotesize\caption{DWM-RO architecture.
\textbf{Module 1:} Each agent is equipped with a world model, which consists of two learning phases. First, an RSSM is trained on real environmental interactions to learn a predictive model of the network dynamics. Second, an AC policy is trained with both real and imagined samples.
\textbf{Module 2:} A dedicated gate operates on each agent. It uses local interference and model reconstruction error to make a binary decision $\mathbb{D}_k(t)$.
\textbf{Module 3:} When offloading is triggered, the agent's latent state $\mathbf{z}_k(t)$ is transmitted to the FBS. It aggregates latents from multiple agents, computes their common component, and returns a decorrelated latent state $\tilde{\mathbf{z}}_k(t)$ to the agent for coordinated action generation.
}
\label{fig:framwork}
\end{figure*}

\section{DWM-RO Framework Design}
As shown in Fig.~\ref{fig:framwork}, DWM-RO employs a hierarchical architecture where each FBS-FUE pair operates as an autonomous MARL agent, with selective coordination facilitated by the FBS. The framework comprises three synergistic modules:

{
\begin{itemize}
    \item \textbf{Decentralized World Models}: Each FBS-FUE$_k$ pair maintains a dedicated recurrent state-space model (RSSM) \cite{hafner2025nature} that learns the temporal dynamics of channels, interference patterns, and EH opportunities, allowing model-based planning for joint beamforming and power splitting decisions.
    
    \item \textbf{Uncertainty-Aware Offloading}: A gating mechanism monitors prediction confidence and environmental volatility, adaptively triggering edge assistance when agents' local world models face high uncertainty from inter-link interference or rapid channel variations.
    
   \item \textbf{Edge-Assisted Latent Refinement}: A lightweight coordination mechanism as the edge-side component of Reasoning Offloading at the FBS refines latent representations from offloading agents to resolve inter-link conflicts. By identifying and removing shared strategic intent, it enhances system-wide performance with minimal computational overhead.
\end{itemize}
}
This hierarchical design achieves computational tractability through distributed execution while preserving coordination benefits when environmental complexity exceeds local modeling capabilities. The following subsections detail the technical design of each module.

\subsection{Decentralized World Model}

Each FBS-FUE$_k$ pair operates as an autonomous agent built upon a world model that learns a predictive representation of the environment dynamics. The world model consists of an RSSM for learning temporal dynamics, along with three predictive components: a dynamics predictor, an observation decoder, and a reward predictor. This architecture enables an agent to perform imagination-based planning by simulating future trajectories entirely within a learned latent space, dramatically improving sample efficiency without requiring extensive real environment interactions.

As shown in Fig.~\ref{fig:framwork}, in time slot $t$, the RSSM compresses the high-dimensional observation $\mathbf{o}_k(t)$ into a compact latent state representation composed of a deterministic state $\mathbf{d}_k(t)$ for temporal memory and a stochastic state $\mathbf{z}_k(t)$ to capture uncertainty. First, the evolution of $\mathbf{d}_k(t)$ follows a deterministic transition, i.e.,
\begin{equation}
    \mathbf{d}_k(t) = f_\phi(\mathbf{d}_k(t-1), \mathbf{z}_k(t-1), \mathbf{a}_k(t-1)), \label{eq:det_trans}
\end{equation}
where $f_\phi$ denotes a Recurrent Neural Network (RNN)~\cite{sherstinsky2020fundamentals,hafner2025nature}, which captures long-term temporal dependencies in the sequential observations and actions. The action $\mathbf{a}_k(t-1) \in \mathbb{R}^{2N_F+1}$ represents the action in the time slot $t-1$. The stochastic state $\mathbf{z}_k(t-1)$ is generated by an observation encoder, which is defined as
\begin{align} 
   \quad &\mathbf{z}_k(t) \sim q_\phi(\mathbf{z}_k \mid \mathbf{d}_k(t), \mathbf{o}_k(t)), \label{eq:encoder}
\end{align}
where $q_\phi$ is a posterior network (implemented as a multi-layer perceptron) that infers the stochastic state from the current observation $\mathbf{o}_k(t)$ and the deterministic state $\mathbf{d}_k(t)$.

In addition to the RSSM core, the world model includes three predictive components for imagination-based planning, such as
\begin{align}
    \text{Dynamics predictor:} \quad &\hat{\mathbf{z}}_k(t) \sim p_\phi(\hat{\mathbf{z}}_k \mid \mathbf{d}_k(t)), \label{eq:prior}\\
    \text{Observation decoder:} \quad &\hat{\mathbf{o}}_k(t) \sim p_\phi(\hat{\mathbf{o}}_k \mid \mathbf{d}_k(t), \mathbf{z}_k(t)), \label{eq:decoder}\\
    \text{Reward predictor:} \quad & \hat{r}_k(t) \sim p_\phi(\hat{r}_k \mid \mathbf{d}_k(t), \mathbf{z}_k(t)). \label{eq:reward_pred}
\end{align}
Note that $\mathbf{d}_k(t)$ is acquired by~\eqref{eq:det_trans}. Therefore, we can observe that the dynamics predictor $p_\phi(\mathbf{z}_k \mid \mathbf{d}_k(t))$ predicts future stochastic states without access to future observations, enabling planning through imagination~\cite{hafner2019dream}. The observation decoder reconstructs the high-dimensional observation from the latent state, ensuring information preservation. The reward predictor estimates expected rewards from latent states~\cite{hafner2025nature,worldmodel2}. All these components share the same parameter set $\phi$ and are trained jointly with the RSSM on collected experience data. The corresponding process is illustrated in Fig.~\ref{fig:framwork}.

{
The agent interacts with the communication environment through a state-action-reward framework designed to directly solve the optimization problem $\mathcal{P}_1$. The agent's observation consists of locally measurable quantities, which are defined as
\begin{equation}
\mathbf{o}_k(t) = \{\mathbf{h}_k(t), I_k^{\text{total}}(t), E_k(t-1), R_k^{\text{FUE}}(t-1), \mathcal{V}_k(t-1)\},
\end{equation}
where $\mathbf{h}_k(t) \in \mathbb{C}^{N_F \times 1}$ is the local channel state vector from the FBS to FUE $k$ in~\eqref{eq:fue_signal}, $I_k^{\text{total}}(t) = I_k^{\text{co}}(t) + I_k^{\text{sat}}(t)$ represents the measured total interference, $E_k(t-1)$ and $R_k^{\text{FUE}}(t-1)$ are the previous harvested power and achieved rate, and $\mathcal{V}_k(t-1) \in \{0,1\}^3$ indicates constraint violations from the previous time step.

The observation features are selected according to four principles: local measurability, direct relevance to $\mathcal{P}_1$, predictive usefulness for the RSSM, and compact dimensionality. In particular, $\mathbf{h}_k(t)$ captures the instantaneous serving-link quality, while $I_k^{\text{total}}(t)$ summarizes the local interference condition that drives both uncertainty and coordination demand. The previous harvested power $E_k(t-1)$ and achieved rate $R_k^{\text{FUE}}(t-1)$ provide compact summaries of recent task outcomes, and $\mathcal{V}_k(t-1)$ explicitly indicates whether recent actions violated QoS or EH constraints. By contrast, global CSI, other agents' observations, and full cross-tier channel matrices are not included because they are either unavailable under decentralized execution or unnecessarily enlarge the input space. This design keeps the observation locally accessible while remaining sufficiently informative for latent prediction and control.
}
The agent employs an actor network $\pi_k$ that maps the latent state to actions by
\begin{align}
    \mathbf{a}_k(t) =& \pi_k(\mathbf{d}_k(t), \mathbf{z}_k(t)) \nonumber\\
    =& [\text{Re}(\mathbf{w}_k), \text{Im}(\mathbf{w}_k), \alpha_k^{\text{raw}}] \in \mathbb{R}^{2N_F+1}, \label{worldmodelactor}
\end{align}
where the first $2N_F$ elements parameterize the complex beamforming vector $\mathbf{w}_k$ and the last element determines the power splitting ratio $\alpha_k$ after sigmoid transformation.

The reward function guides the agent toward solving $\mathcal{P}_1$ by combining the primary objective from~\eqref{eq:obj} with penalties for constraint violations which is defined as
\begin{equation}
    r_k(t) = \omega \cdot R_k^{\text{FUE}}(t) - \mathcal{P}_k(t), \label{eq:main_reward}
\end{equation}
where $\omega$ is a scaling factor, and $\mathcal{P}_k(t)$ aggregates constraint violations given by
\begin{align}
    \mathcal{P}_k(t) = &\lambda_1 \cdot \text{ReLU}(\xi_{\text{FUE}} - R_k^{\text{FUE}}(t)) \label{eq:penalty_fue_qos} \\
    &+ \lambda_2 \cdot \text{ReLU}(\Phi_{\text{FUE}} - E_k(t)) \label{eq:penalty_eh} \\
    &+ \lambda_3 \sum_{m \in \mathcal{M}} \text{ReLU}(\xi_{\text{SUE}} - R_m^{\text{SUE}}(t)) \cdot w_{k,m}(t), \label{eq:penalty_sue_qos}
\end{align}
where $\lambda_1$, $\lambda_2$, and $\lambda_3$ are penalty weights, and $w_{k,m}(t) = I_{k,m}(t)/\sum_{j} I_{j,m}(t)$ attributes cross-tier interference responsibility based on agent $k$'s contribution to SUE $m$'s total interference.

{
The reward structure in~\eqref{eq:main_reward} incorporates a Lagrangian relaxation of $\mathcal{P}_1$, where $\omega$ normalizes the spectral efficiency objective and $\lambda_1, \lambda_2, \lambda_3$ serve as penalty multipliers that drive feasibility. In practice, $\omega$ is first calibrated so that the rate reward is of comparable magnitude to the penalty terms, after which the penalty weights are increased until the trained policy consistently satisfies all constraints. Since the ReLU-based penalty terms in~\eqref{eq:penalty_fue_qos} to~\eqref{eq:penalty_sue_qos} are identically zero whenever the constraints are met, the framework is robust to moderate variations in these weights, and the primary requirement is that the penalty-to-objective ratio is large enough to enforce feasibility at the constraint boundary.
}

The world model is trained on collected experience by minimizing the variational free energy objective. For notational clarity, we omit the agent index $k$ in the following loss functions and use $\mathbf{d}_t$ and $\mathbf{z}_t$ instead of $\mathbf{d}_k(t)$ and $\mathbf{z}_k(t)$. The loss of RSSM is defined as 
\begin{equation}
    \mathcal{L}_{\text{RSSM}}^k = \mathbb{E}_{q_\phi}\left[\sum_{t=1}^{T} \mathcal{L}_{\text{pred}}(t) + \beta_{\text{dyn}}\mathcal{L}_{\text{dyn}}(t) + \beta_{\text{rep}}\mathcal{L}_{\text{rep}}(t)\right],
\end{equation}
where:
\begin{align}
    \mathcal{L}_{\text{pred}}(t) &= -\log p_\phi(\hat{\mathbf{o}}_t | \mathbf{d}_t, \mathbf{z}_t) - \log p_\phi(\hat{r}_t | \mathbf{d}_t, \mathbf{z}_t), \\
    \mathcal{L}_{\text{dyn}}(t) &= \max(1, \text{KL}[\text{sg}(q_\phi(\mathbf{z}_t | \mathbf{d}_t, \mathbf{o}_t)) \| p_\phi(\hat{\mathbf{z}}_t | \mathbf{d}_t)]), \\
    \mathcal{L}_{\text{rep}}(t) &= \max(1, \text{KL}[q_\phi(\mathbf{z}_t | \mathbf{d}_t, \mathbf{o}_t) \| \text{sg}(p_\phi(\hat{\mathbf{z}}_t | \mathbf{d}_t))]),
\end{align}
with $\beta_{\text{dyn}} = 1$ and $\beta_{\text{rep}} = 0.1$. $\text{sg}(\cdot)$ denotes the stop-gradient operator. The prediction loss $\mathcal{L}_{\text{pred}}$ ensures accurate reconstruction and reward prediction; the dynamics loss $\mathcal{L}_{\text{dyn}}$ trains the sequence model to predict the next representation by minimizing the Kullback-Leibler (KL) divergence between the posterior and the predictor, and the representation loss $\mathcal{L}_{\text{rep}}$ encourages predictable latent states. The free bits mechanism, implemented via the $\max(1, \cdot)$ operation, clips both KL losses below 1 nat to avoid degenerate solutions and enable fixed hyperparameters across domains.

Once the world model has been trained, the actor-critic networks learn through imagination rollouts. Starting from a real latent state $(\mathbf{d}_t, \mathbf{z}_t)$, the model generates trajectories of length $H$ entirely within the latent space as
\begin{align}
    \hat{\mathbf{d}}_{i+1} &= f_\phi(\hat{\mathbf{d}}_i, \hat{\mathbf{z}}_i, \hat{\mathbf{a}}_i), \quad i = t, \ldots, t+H-1, \\
    \hat{\mathbf{z}}_{i+1} &\sim p_\phi(\mathbf{z} | \hat{\mathbf{d}}_{i+1}), \\
    \hat{r}_{i+1} &\sim p_\phi(r | \hat{\mathbf{d}}_{i+1}, \hat{\mathbf{z}}_{i+1}),
\end{align}
where $\hat{\mathbf{a}}_i \sim \pi_k(\hat{\mathbf{d}}_i, \hat{\mathbf{z}}_i)$ are actions sampled from the current policy. The critic $V_k(\mathbf{d}_t, \mathbf{z}_t)$ learns to estimate expected returns from these imagined trajectories. The loss of the critic $\mathcal{L}_{\text{critic}}^k$ is defined as
\begin{equation}
    \mathcal{L}_{\text{critic}}^k = \mathbb{E}_{\hat{\tau}}\left[\sum_{i=t}^{t+H} \left(V_k(\hat{\mathbf{d}}_i, \hat{\mathbf{z}}_i) - R_i^\lambda\right)^2\right],
\end{equation}
where $R_i^\lambda$ is the $\lambda$-return calculated on the imagined trajectory. The actor maximizes expected imagined returns by minimizing the loss that is defined as
\begin{equation}
    \mathcal{L}_{\text{actor}}^k = -\mathbb{E}_{\hat{\tau}}\left[\sum_{i=t}^{t+H} \hat{r}_i\right].
\end{equation}

This separation between world model training on real data and policy training on imagined data addresses \textbf{Challenge I} by enabling policy optimization with reduced environment interactions. Each real experience can generate multiple imagined rollouts.

\subsection{Uncertainty-Aware Offloading}
Although decentralized world models enable autonomous operation, we observe that certain scenarios exceed the predictive capabilities of local agents. High interference levels and degraded model predictions can create situations where local decision-making is convergent. Hence, we introduce an uncertainty-aware offloading gate that adaptively determines when edge-assisted decorrelation is beneficial.

The offloading gate operates as a binary decision module at each FUE $k$. In time slot $t$, it decides whether to act locally with the original latent state $\mathbf{z}_k(t)$ or to offload this latent to the edge server for decorrelation. When offloading is triggered, the edge server applies the latent decorrelation mechanism (detailed in Section IV.C) to produce a decorrelated latent $\tilde{\mathbf{z}}_k(t)$. The agent then generates its action, i.e.,
\begin{equation}
\mathbf{a}_k(t) = 
\begin{cases}
\pi_k(\mathbf{d}_k(t), \mathbf{z}_k(t)), & \text{if } \mathbb{D}_k(t) = 0 \text{ (local)}, \\
\pi_k(\mathbf{d}_k(t), \tilde{\mathbf{z}}_k(t)), & \text{if } \mathbb{D}_k(t) = 1 \text{ (decorrelated)}.
\end{cases}
\end{equation}

The gate bases its decision on uncertainty indicators which are defined as
\begin{equation}
\mathbf{u}_k(t) = [I_k^{\text{total}}(t), \mathcal{E}_k^{\text{rec}}(t)] \in \mathbb{R}^2,
\label{eq:gate_input}
\end{equation}
where $I_k^{\text{total}}(t) = I_k^{\text{co}}(t) + I_k^{\text{sat}}(t)$ represents the total interference level, and $\mathcal{E}_k^{\text{rec}}(t) = \|\hat{\mathbf{o}}_k(t) - \mathbf{o}_k(t)\|_2^2$ is the reconstruction error from the world model's observation decoder.

The gate employs a dedicated actor-critic architecture given by
\begin{itemize}
    \item \textbf{Gate-Actor} $\pi_k^{\text{gate}}(\cdot|\mathbf{u}_k; \xi_k^{\text{actor}})$: A two-layer Multilayer perceptron (MLP) with hidden dimension 64 that outputs the offloading probability.
    \item \textbf{Gate-Critic} $V_k^{\text{gate}}(\mathbf{u}_k; \xi_k^{\text{critic}})$: Estimates the expected return for state $\mathbf{u}_k$.
\end{itemize}

To train the gate, we design an asymmetric reward structure that accounts for the availability of information:

\begin{itemize}
    \item \textbf{Offloading} ($\mathbb{D}_k(t) = 1$): The edge server returns $\tilde{\mathbf{z}}_k(t)$, and the agent acts using $\pi_k(\mathbf{d}_k(t), \tilde{\mathbf{z}}_k(t))$, obtaining real reward $r_k^{\text{dec}}(t)$. Since we have access to both the original and decorrelated latents, we can estimate what would have happened without decorrelation: $\hat{r}_k^{\text{local}}(t) = \mathbb{E}_{p_\phi}[r | \mathbf{d}_k(t), \mathbf{z}_k(t)]$.
    
    \item \textbf{Local Execution} ($\mathbb{D}_k(t) = 0$): The agent acts directly using $\pi_k(\mathbf{d}_k(t), \mathbf{z}_k(t))$ and obtains real reward $r_k^{\text{local}}(t)$. Since no offloading occurred, we have no information about what the decorrelated latent would have been.
\end{itemize}

Given this asymmetry, we define the gate's reward signal as
\begin{equation}
\mathcal{R}_k^{\text{gate}}(t) = 
\begin{cases}
r_k^{\text{local}}(t), & \text{if } \mathbb{D}_k(t) = 0, \\
r_k^{\text{dec}}(t) - c, & \text{if } \mathbb{D}_k(t) = 1,
\end{cases}
\label{eq:gate_reward}
\end{equation}
where $c > 0$ is a communication cost that penalizes offloading overhead.

When offloading occurs, we can additionally compute the improvement metric $\Delta_k(t) = r_k^{\text{dec}}(t) - \hat{r}_k^{\text{local}}(t)$ to track whether decorrelation helped. This metric is logged for analysis but not directly used in training, as the gate learns from the actual rewards received.

This formulation encourages the gate to offload when the improvement from decorrelation $(r_k^{\text{dec}}(t) - \hat{r}_k^{\text{local}}(t))$ exceeds the communication cost. When the gate chooses local execution, it receives the actual reward minus the communication cost it saved. Over time, the gate learns to identify situations where decorrelation provides sufficient benefit to justify the overhead, naturally discovering patterns of high interference or correlated agent behaviors where coordination is most valuable.

\subsection{Edge-Side Latent Refinement}
When one or more FUEs trigger their offloading gates, the FBS receives their respective latent states, initiating the coordination phase. The objective at the edge is to refine these latent representations to mitigate inter-agent conflicts and enhance system-wide performance before the agents generate their final actions. Instead of employing computationally expensive architectures (e.g., Transformer \cite{mo2025diffuse,zhang2024decentralized}), we propose a lightweight, efficient, and highly effective mechanism for latent decorrelation.

The process is executed at the FBS and consists of the following steps:
\subsubsection{Latent State Aggregation and Mutual Component Identification} At time $t$, the FBS collects the set of offloaded latent states from the coordinating group of agents $\mathcal{O}(t)$, i.e.,
\begin{equation}
\mathcal{Z}(t)^{\text{offload}} = {\mathbf{z}_{k\in \mathcal{O}(t)}(t)}.
\end{equation}
To identify the shared strategic intent that may lead to conflict, the FBS computes computes the shared latent component of these states as
\begin{equation}
\bar{\mathbf{z}}(t) = \frac{1}{|\mathcal{O}(t)|} \sum_{k \in \mathcal{O}(t)} \mathbf{z}_k(t).
\label{eq:mutual_component}
\end{equation}
The resulting vector $\bar{\mathbf{z}}(t)$ represents the average belief of the group. Its high-magnitude dimensions signify strategic intentions common across the agents (e.g., a shared tendency to select high transmission power), which are likely sources of resource contention.
\subsubsection{Latent Refinement via Decorrelation}
The core of our coordination mechanism is to foster strategic diversity by removing this shared mutual component from each agent's individual latent state. This operation encourages each agent to act upon the unique aspects of its own belief state, thereby reducing strategic overlap. The refined, or ``decorrelated," latent state for each agent $k \in \mathcal{O}(t)$ is computed as
\begin{equation}
\tilde{\mathbf{z}}_k(t) = \mathbf{z}_k(t) - \bar{\mathbf{z}}(t).
\label{eq:latent_refinement}
\end{equation}

{
\begin{proposition}
Let $\boldsymbol{\delta}_k(t) = \mathbf{z}_k(t) - \bar{\mathbf{z}}(t)$ denote the refined latent state for agent $k \in \mathcal{O}(t)$, where $\bar{\mathbf{z}}(t)$ is the group mean defined in~\eqref{eq:mutual_component}. Then, the aggregate pairwise inner product satisfies
\begin{equation}
    \sum_{\substack{i, j \in \mathcal{O}(t) \\ i \neq j}} \boldsymbol{\delta}_i(t)^\top \boldsymbol{\delta}_j(t) = -\sum_{k \in \mathcal{O}(t)} \left\|\boldsymbol{\delta}_k(t)\right\|^2 \leq 0.
\end{equation}
\end{proposition}
}
{
\begin{IEEEproof}
By the definition of the group mean, we have $\sum_{k \in \mathcal{O}(t)} \boldsymbol{\delta}_k(t) = \mathbf{0}$. Expanding the squared norm of this sum yields
\begin{align}
    \mathbf{0} &= \left\|\sum_{k} \boldsymbol{\delta}_k(t)\right\|^2 = \sum_{k} \left\|\boldsymbol{\delta}_k(t)\right\|^2 + \sum_{\substack{i \neq j}} \boldsymbol{\delta}_i(t)^\top \boldsymbol{\delta}_j(t).
\end{align}
Rearranging completes the proof.
\end{IEEEproof}

This subtraction ensures that the refined latent state $\tilde{\mathbf{z}}_k(t)$ emphasizes the unique, distinguishing features of each agent's strategy.
}
\subsubsection{Coordinated Action Generation} 
The FBS transmits the refined latent states ${\tilde{\mathbf{z}}_k(t)}_{k \in \mathcal{O}(t)}$ back to the corresponding FUE $k$. Crucially, the edge server does not dictate actions. Instead, each agent uses its own local policy network $\pi_k$ to generate a new, coordinated action from the refined latent as
\begin{equation}
\tilde{\mathbf{a}}_k(t) = \pi_k(\mathbf{d}_k(t), \tilde{\mathbf{z}}_k(t))
\end{equation}
This allows the agents to utilize and benefit from centralized coordination while retaining their specialized, locally learned policies. The resulting actions $\tilde{\mathbf{w}}_k(t)$ and $\tilde{\alpha}_k(t)$ are better coordinated, leading to reduced interference and a more efficient allocation of shared resources.

{
This lightweight edge-side mechanism is highly scalable, with computational complexity that grows only linearly with the number of offloading agents, making it ideal for resource-limited wireless communication systems. A conceptual illustration is provided in Fig.~\ref{fig:decorrelation_effect}.

\begin{figure*}[tpb]
    \centering
    \includegraphics[width=\textwidth]{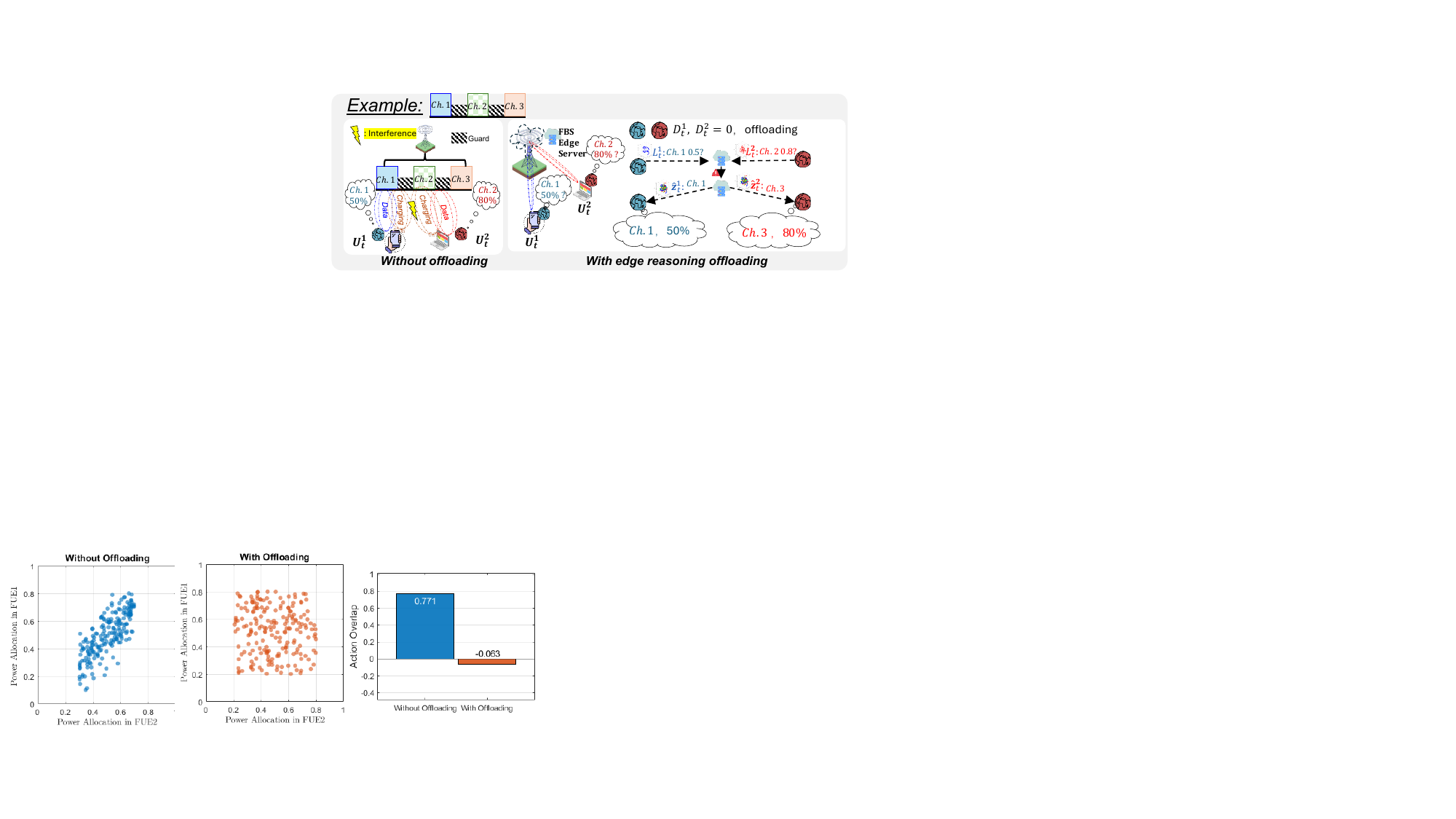} 
    \caption{\footnotesize Conceptual illustration of the uncertainty-aware reasoning offloading mechanism. Without coordination (left), agents independently select conflicting channels, causing mutual interference. With edge reasoning offloading (right), agents offload their latent states to the FBS, which removes the shared latent component and returns decorrelated representations, guiding agents toward complementary resource allocation.}
    \vspace{-0.3cm}
    \label{fig:decorrelation_effect}
\end{figure*}

\begin{example} 
    Consider two FUEs that, without coordination, both select Channel~1 with similar power splitting ratios, resulting in severe mutual interference~\cite{8930071}. When the offloading gate is activated, both agents transmit their latent states $\mathbf{z}_1(t)$ and $\mathbf{z}_2(t)$ to the FBS. The edge server identifies and removes the shared latent component $\bar{\mathbf{z}}(t)$, yielding agent-specific latent components that emphasize each agent's unique strategic information. As a result, the agents are guided toward non-interfering channels with complementary power allocations, resolving the resource conflict.
\end{example}

The quantitative impact of this coordination mechanism on both the latent representation space and the action space is presented in Section~\ref{sec:latent_analysis}, where we show that the shared component removal reduces action overlap from $0.771$ to $-0.063$, directly addressing \textbf{Challenge~II} regarding the lack of effective coordination among decentralized agents.
}

\begin{algorithm}[tpb]
\caption{Distributed Execution Protocol}
\label{alg:execution}
\begin{algorithmic}[1]
\STATE \textbf{Phase 1: Belief Update and Gate Decision}
\FOR{each agent $k \in \{1, \ldots, K\}$ in parallel}
    \STATE Update belief: $(\mathbf{d}_k(t), \mathbf{z}_k(t))$ using RSSM
    \STATE Compute gate input: $\mathbf{u}_k(t) = [I_k^{\text{total}}(t), \mathcal{E}_k^{\text{rec}}(t)]$
    \STATE Sample gate decision: $\mathbb{D}_k(t) \sim \pi_k^{\text{gate}}(\cdot | \mathbf{u}_k(t))$
\ENDFOR

\STATE \textbf{Phase 2: Edge Refinement (if triggered)}
\IF{$\mathcal{O}(t) = \{k : \mathbb{D}_k(t) = 1\} \neq \emptyset$}
    \STATE Apply latent decorrelation on $\{\mathbf{z}_k(t)\}_{k \in \mathcal{O}(t)}$
    \STATE Obtain refined latents $\{\tilde{\mathbf{z}}_k(t)\}_{k \in \mathcal{O}(t)}$
\ENDIF

\STATE \textbf{Phase 3: Action Generation}
\FOR{each agent $k \in \{1, \ldots, K\}$ in parallel}
    \IF{$\mathbb{D}_k(t) = 1$}
        \STATE $\mathbf{a}_k(t) = \pi_k(\mathbf{d}_k(t), \tilde{\mathbf{z}}_k(t))$
    \ELSE
        \STATE $\mathbf{a}_k(t) = \pi_k(\mathbf{d}_k(t), \mathbf{z}_k(t))$
    \ENDIF
    \STATE Extract and normalize beamforming and PS ratio
\ENDFOR

\STATE \textbf{Return:} $\{(\mathbf{w}_k(t), \alpha_k(t))\}_{k=1}^K$
\end{algorithmic}
\end{algorithm}
\subsection{Distributed Training and Execution Protocol}
{ 
The proposed hierarchical framework is realized through a fully distributed protocol for both training and execution. During the training phase, each of the $K$ agents concurrently and independently learns its respective network components. The world model (RSSM), parameterized by $\phi_k$, is updated using collected environmental transitions by minimizing the variational objective $\mathcal{L}_{\text{RSSM}}^k$. Based on the learned latent dynamics, the main actor-critic policy is then refined using ``imagination rollouts,'' which are trajectories generated entirely within the latent space of this world model to enhance sample efficiency. In parallel, the offloading gate is optimized from task returns under local and edge-assisted execution, enabling each agent to adaptively decide when coordination should be invoked.

Once trained, the system runs in real time according to the execution protocol detailed in Algorithm~\ref{alg:execution}. At each time slot $t$, an agent first updates its internal belief state $(\mathbf{d}_k(t), \mathbf{z}_k(t))$, decides whether to offload via its gate, and then generates a raw action $\mathbf{a}_k(t)$. This raw action is subsequently post-processed to derive the final, physically compliant control parameters. The beamforming vector $\mathbf{w}_k(t)$ is normalized to satisfy the power constraint~\eqref{eq:power}, i.e.,
\begin{equation}
    \mathbf{w}_k(t) = \sqrt{P_{\max}} \cdot \frac{\mathbf{a}_k(t)[1:2N_F]}{\|\mathbf{a}_k(t)[1:2N_F]\|},
\end{equation}
where $\mathbf{a}_k(t)[1:2N_F]$ represents the complex vector formed from the raw action. The power splitting ratio $\alpha_k(t)$ is passed through a sigmoid function to ensure that it is within $[0, 1]$, satisfying constraint~\eqref{eq:ps_range}. $\alpha_k(t)$ is defined as
\begin{equation}
    \alpha_k(t) = \sigma(\mathbf{a}_k(t)[2N_F+1]),
\end{equation}
where $\sigma(\cdot)$ is the sigmoid function. The offloading round-trip is designed to complete well within a single time slot. The transmitted payload consists of a compact 32-dimensional latent vector (256 bytes total), which transfers in approximately 0.2~ms at typical femtocell rates. Over this interval, the Jakes channel correlation remains $\rho = J_0(2\pi f_d \Delta t) \approx 0.99996$ for $f_d = 10$~Hz (TABLE~\ref{tab:simulation_parameters}), indicating negligible channel variation. The framework is further robust to such sub-millisecond delays because the deterministic recurrent state $\mathbf{d}_k(t)$, which encodes the full temporal history, is computed locally before any communication occurs and is never transmitted. The gate mechanism adaptively manages the latency-performance trade-off through the communication cost penalty $c$ in~\eqref{eq:gate_reward}, ensuring that offloading is triggered only when the expected coordination benefit outweighs the overhead. This complete protocol enables agents to operate autonomously while selectively using edge coordination, achieving effective and adaptive resource allocation in the SWIPT-enabled HetNet.
}

\section{Numerical Results}
\begin{table}[tpb]
\centering
\caption{Consolidated Simulation Parameters}
\label{tab:simulation_parameters}
\footnotesize 
 
\begin{tabular}{@{}lll@{}}
\toprule
\textbf{Category} & \textbf{Parameter} & \textbf{Value} \\ \midrule
\multicolumn{3}{l}{\textit{\textbf{A. Network Topology and Scale}}} \\
Number of SUEs & $M$ & 3 \\
Number of FUEs & $K$ & 2 \\
FBS Antennas & $N_F$ & 6 \\
Satellite Antennas & $N_M$ & 8 \\ 
\cmidrule(l){1-3} 
\multicolumn{3}{l}{\textit{\textbf{B. Channel, Power, and Noise}}} \\
Doppler Frequency & $f_d$ & 10 Hz \\
Time Slot Duration & $T_s$ & 0.001 s \\
Rician Factor & $K_{\text{rician}}$ & 4 \\
FBS Max Transmit Power & $P_{\max}$ (Constr.~\eqref{eq:power}) & 20 dBm \\
Satellite Transmit Power & $P_{\text{sat}}$ & 43 dBm\\
SUE Noise Power & $\sigma_a^2$ & -70 dBm \\
FUE Noise Power & $\sigma_b^2$ & -75 dBm \\ 
\cmidrule(l){1-3} 
\multicolumn{3}{l}{\textit{\textbf{C. QoS and EH Constraints}}} \\
SUE Min. Rate & $\xi_{SUE}$ (Constr.~\eqref{eq:sue_qos}) & 0.5 bps/Hz \\
FUE Min. Rate & $\xi_{FUE}$ (Constr.~\eqref{eq:fue_qos}) & 0.3 bps/Hz \\
FUE Min. EH & $\Phi_{FUE}$ (Constr.~\eqref{eq:eh_constraint}) & 0.1 mW \\
Max Harvestable Power & $E_{\max}$ (Eq.~\eqref{eq:EH_nl}) & 24 mW \\
EH Circuit Sensitivity & $\mu$ (Eq.~\eqref{eq:EH_nl}) & 150 \\
EH Turn-on Threshold & $\nu$ (Eq.~\eqref{eq:EH_nl}) & 0.024 mW \\ 
\cmidrule(l){1-3} 
\multicolumn{3}{l}{\textit{\textbf{D. DRL Agent Training}}} \\
World Model Learning Rate & $\eta_{wm}$ & $6 \times 10^{-4}$ \\
Actor-Critic Learning Rates & $\eta_{actor}, \eta_{critic}$ & $3 \times 10^{-4}$ \\
Discount Factor & $\gamma$ & 0.99 \\
Training Batch Size & - & 16 \\
Number of Episodes & - & 20,000 \\
Episode Length & - & 20 steps \\
Imagination Horizon & $H$ & 5 \\ \bottomrule
\end{tabular}
\end{table}
\subsection{Experimental Setup}
The simulation environment is designed to instantiate the SWIPT-enabled HetNet system model described in Section~\ref{sec:system_model}. This includes the time-varying Jakes fading channel~\cite{xiao2002second}, as described in~\eqref{eq:jakes_fading} and \eqref{eq:jakes_rho}, and the practical non-linear EH model from~\eqref{eq:EH_nl}. All key parameters for the testbed and MARL training are consolidated in TABLE~\ref{tab:simulation_parameters}.

\subsubsection{Implementation Details}
Our framework is implemented in PyTorch. As detailed in Section IV, each FBS-FUE$_k$ pair is an autonomous agent based on a comprehensive world model architecture. The agent's action $\mathbf{a}_k(t)$ is a continuous vector of dimension $2N_F+1=13$, directly parameterizing the beamforming vector $\mathbf{w}_k(t)$ and power splitting ratio $\alpha_k(t)$. The agent's learning process is guided by the reward function defined in~\eqref{eq:main_reward}, which directly translates the constrained optimization problem $\mathcal{P}_1$ into a learning objective. The RSSM at the core of each agent has a 256-dimensional deterministic state and a 32-dimensional stochastic state. The actor and critic networks are 2-layer MLPs with 256 hidden units and are trained using the Adam optimizer~\cite{Kingma2014AdamAM}.

\subsubsection{Evaluation Metrics}
We assess the performance of all methods using three key metrics that directly correspond to the objective and constraints of the optimization problem $\mathcal{P}_1$.
\begin{itemize}
    \item \textbf{Average Sum-Rate of FUEs:} The time-averaged sum of the achievable rates of all FUEs, i.e., $\sum_{k \in \mathcal{K}} R_k^{\text{FUE}}(t)$, measured in bps/Hz. This directly evaluates performance against the primary objective~\eqref{eq:obj}.
    
    \item \textbf{Constraint Violation Rate:} The percentage of time steps during which one or more of SUE QoS~\eqref{eq:sue_qos}, FUE QoS~\eqref{eq:fue_qos}, and FUE EH~\eqref{eq:eh_constraint} constraints is not satisfied. This quantifies the reliability of a solution.
    
    \item \textbf{Average Harvested Power:} The time-averaged power harvested by each FUE, which evaluates the effectiveness in satisfying the EH constraint~\eqref{eq:eh_constraint}.
\end{itemize}


\subsection{Performance of the Base World Model}

{
First, we compare Pure DWM against a comprehensive set of baselines. Pure DWM serves as the model-based ablation baseline of DWM-RO. It preserves the same RSSM-based predictive backbone and imagination-based actor-critic training, while removing the uncertainty-aware offloading gate and the edge-side latent refinement mechanism. Therefore, the comparison between DWM-RO and Pure DWM isolates the contribution of reasoning offloading beyond model-based latent planning itself. This set includes state-of-the-art model-free reinforcement learning algorithms, namely Proximal Policy Optimization (PPO)~\cite{gu2021proximal}, Soft Actor-Critic (SAC)~\cite{haarnoja2018soft}, and Deep Q-Network (DQN)~\cite{fan2020theoretical}. Additionally, we include a non-learning heuristic EGT (Equal Gain Transmission) policy and a Random policy to serve as lower-bound references.
}

\begin{figure*}[t!]
    \centering
    \includegraphics[width=0.9\linewidth]{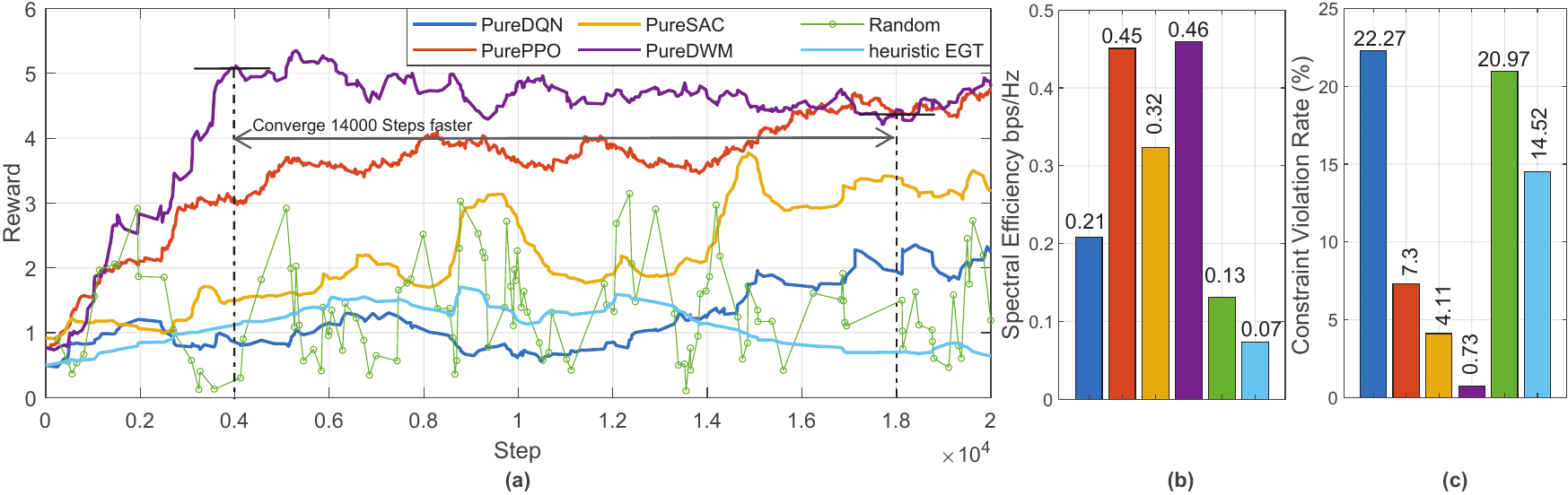} 
    \vspace{-0.3cm}
    \caption{\footnotesize Performance evaluation of the base world model (Pure DWM) against baselines without offloading. The proposed Pure DWM demonstrates superior convergence speed, final reward, spectral efficiency, and constraint satisfaction.}
    \vspace{-0.4cm}
    \label{fig:exp1_baselines}
\end{figure*}

{
The training reward curves in Fig.~\ref{fig:exp1_baselines}\text{(a)} highlight significant differences in learning efficiency, where convergence speed is measured by the number of training steps required to reach the target reward plateau rather than by wall-clock runtime. The Pure DWM agent exhibits the best learning trajectory, achieving a high-reward plateau of approximately 5 around step $0.4 \times 10^4$. In contrast, Pure PPO displays a much slower learning process, only beginning to approach a similar reward level near step $1.8 \times 10^4$. This superior sample efficiency is a direct consequence of the world model's architecture. By learning a predictive model of the environment's dynamics, the agent can perform numerous ``imagination rollouts''~\cite{hafner2019dream} in its compact latent space, allowing policy training on vast simulated experience without costly real-world interaction, which model-free methods like Pure PPO require.
}

Other model-free methods fail to learn a competitive policy. SAC shows high instability, likely due to the non-stationarity of the multi-agent environment from an independent off-policy perspective. DQN fails to learn, showing almost no improvement over the random policy. This is primarily because standard DQN requires discretizing our high-dimensional continuous action space, which leads to a combinatorial explosion of actions and significant precision loss, preventing the Q-network from learning a meaningful value function.

Fig.~\ref{fig:exp1_baselines}\text{(b)} quantifies the converged policies' performance. In terms of spectral efficiency, both Pure DWM and Pure PPO learn effective policies, achieving closely matched top-tier results of $0.46~\text{bps/Hz}$ and $0.45~\text{bps/Hz}$, respectively, confirming both can optimize the primary objective. The performance of SAC ($0.32~\text{bps/Hz}$) and DQN ($0.21~\text{bps/Hz}$) lags significantly.

However, reliability is the decisive factor for practical deployment. Fig.~\ref{fig:exp1_baselines}\text{(c)} reveals a clear difference: the Pure DWM achieves a near-perfect violation rate of only $0.73\%$, while Pure PPO's rate is $4.11\%$, more than five times higher. The reason for Pure DWM's superior reliability lies in its predictive planning mechanism. The RSSM includes a reward predictor, and our reward function heavily penalizes constraint violations. During imagination rollouts, the agent simulates long-term trajectories and accumulates predicted rewards, inherently optimizing the policy to be risk-averse and avoid sequences of actions predicted to cause future breaches. In contrast, model-free methods like PPO are more reactive, learning from past and current rewards via an advantage function, and they lack the long-horizon planning capability to consistently foresee and avoid complex scenarios that result in constraint violations.


\subsection{Efficacy of the Reasoning Offloading Mechanism}

To evaluate the uncertainty-aware offloading and edge-side latent refinement mechanism, we conduct a two-stage analysis. First, we perform a focused experiment to isolate and verify the fundamental benefit of the latent refinement process itself. Then, we evaluate the performance of the complete DWM-RO framework with different gate structures against several key benchmarks.

\subsubsection{Analysis of the Latent Refinement Mechanism}
\label{sec:latent_analysis}

\begin{figure}[h!]
    \centering
    \includegraphics[width=0.9\columnwidth]{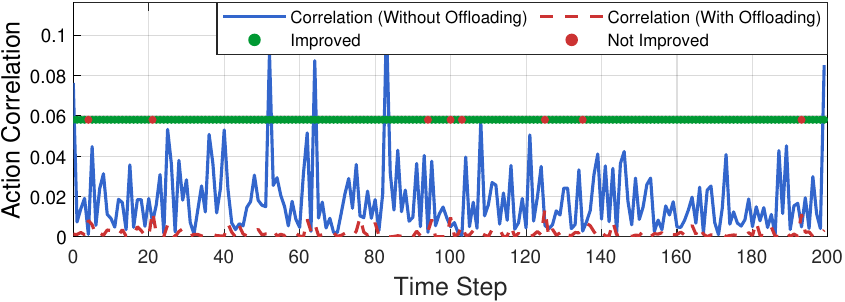}
    \vspace{-0.2cm}
    \caption{\footnotesize Analysis of action correlation and reward improvement over 200 time steps. The ``always offload" strategy dramatically reduces action correlation compared to the baseline. The green dots indicate that this decorrelation results in a reward improvement in nearly every step.}
    \label{fig:decorrelation_timeline}
    \vspace{-0.2cm}
\end{figure}

\begin{figure}[h!]
    \centering
    \includegraphics[width=0.9\columnwidth]{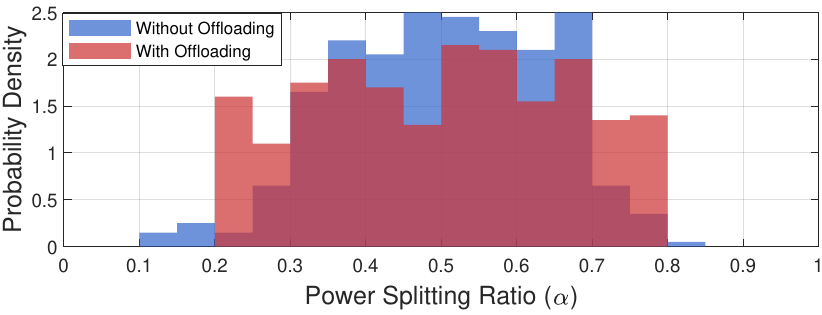}
    \vspace{-0.2cm}
    \caption{\footnotesize Comparison of the probability density of the power splitting ratio ($\alpha$). Agents with ``always offload" explore a visibly different and wider distribution of actions compared to the baseline agents without offloading.}
    \label{fig:distribution_change}
\end{figure}

{

Fig.~\ref{fig:latent_action}(a) and (b) shows the Principal Component Analysis (PCA) projection of agent latent states : before refinement, the two agents' representations overlap significantly in the shared latent space, indicating that both agents encode similar environmental dynamics. After removing the shared latent component via Proposition~1, each agent's cluster separates clearly, confirming that the agent-specific residuals $\boldsymbol{\delta}_k(t)$ capture distinct strategic information as guaranteed by the non-positive pairwise inner product property.

This latent separation directly translates to action diversity. Fig.~\ref{fig:latent_action}(c) and (d) shows the power splitting ratio $\alpha$ of FUE~1 versus FUE~2: without offloading, the agents' decisions are highly correlated (Pearson $\rho = 0.771$), indicating redundant resource allocation. With the offloading mechanism enabled, their actions become effectively decorrelated ($\rho = -0.063$), as summarized in Fig.~\ref{fig:latent_action}(e).

\begin{figure*}[t!]
\centering
\includegraphics[width=0.9\textwidth]{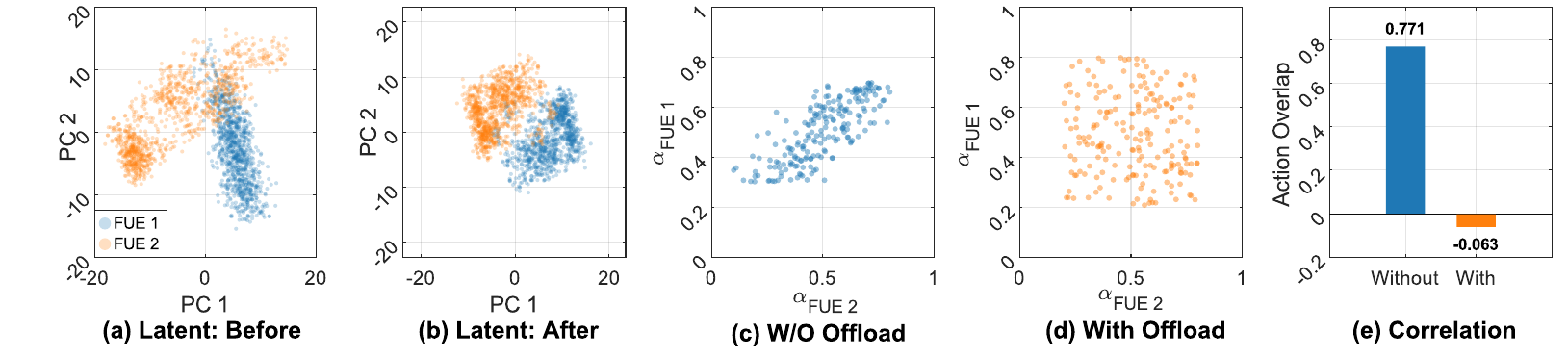}
\caption{\footnotesize Effect of latent decorrelation on agent behavior ($K=2$). (a)(b)~PCA projection of latent states before and after removing the shared latent component: the common environmental encoding is subtracted, yielding clearly separated per-agent clusters. (c)(d)~Power splitting ratio $\alpha$ of FUE~1 vs.\ FUE~2 without and with offloading-based coordination. (e)~Pearson correlation drops from $0.771$ to $-0.063$, confirming that shared component removal induces complementary resource allocation strategies.}
\label{fig:latent_action}
\end{figure*}

We further conduct a 200-step simulation comparing the baseline DWM (Without Offloading) to an ``always offload'' variant. As shown in Fig.~\ref{fig:decorrelation_timeline}, the ``always offload'' strategy successfully suppresses action correlation to near-zero, unlike the fluctuating baseline. This decorrelation fosters behavioral change, as coordinated agents explore a wider distribution of power splitting ratios (Fig.~\ref{fig:distribution_change}) rather than converging on a narrow range. Most importantly, these decorrelated actions translate directly to performance gains, with Fig.~\ref{fig:decorrelation_timeline} showing a reward improvement in $96\%$ of time steps. This confirms the latent refinement mechanism is fundamentally beneficial, motivating the evaluation of our full DWM-RO framework with its adaptive gate.

}

\subsubsection{Performance Comparison of the Full Framework}
\begin{figure*}[t!]
    \centering
    \includegraphics[width=0.9\linewidth]{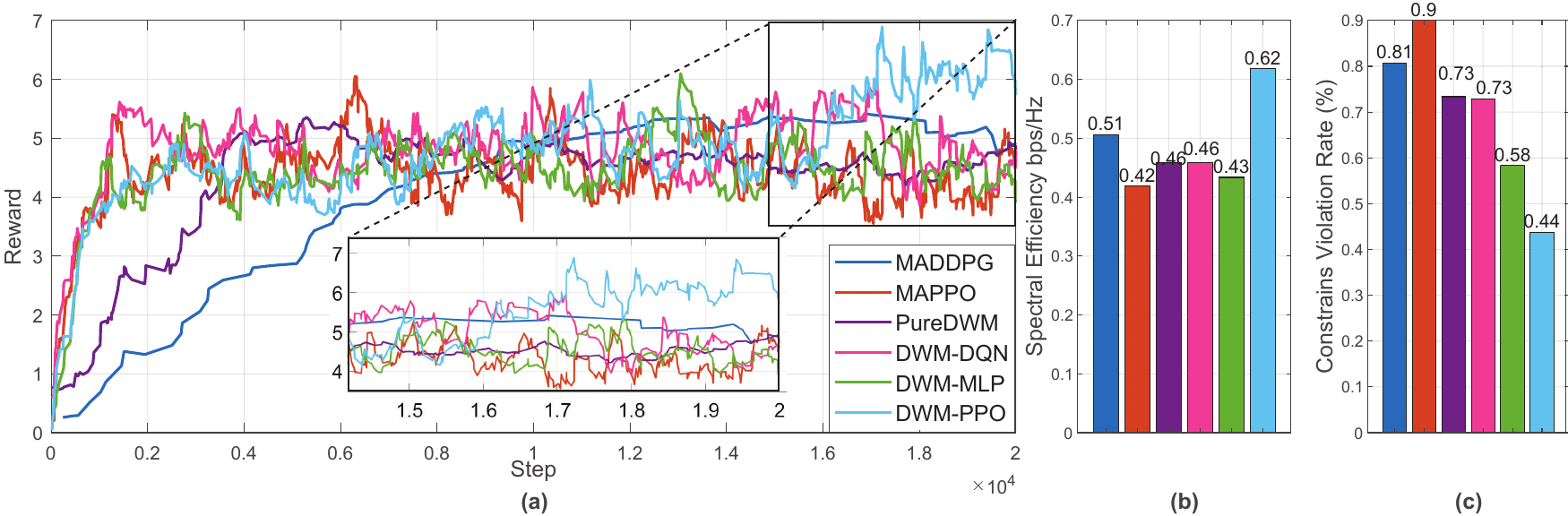} 
    \vspace{-0.3cm}
    \caption{\footnotesize Performance evaluation of the reasoning offloading mechanism. Our full DWM-PPO model is compared against baselines including PureDWM (no offloading), standard MARL algorithms, and ablations of the gate design.}
    \vspace{-0.4cm}
    \label{fig:exp2_offloading}
\end{figure*}

Having established the DWM as a robust base architecture, this experiment evaluates the core contribution of this work: the efficacy of the uncertainty-aware offloading and edge-side latent refinement mechanism. To do so, we compare our full proposed method, DWM-PPO, against several representative baselines shown in Fig.~\ref{fig:exp2_offloading}. These baselines include Pure DWM (with no offloading), conventional MARL algorithms (MAPPO and MADDPG), and ablations of our gate mechanism using a DQN-based gate (DWM-DQN) and an MLP-Bandit (DWM-MLP).

First, we analyze the training dynamics from the reward curves in Fig.~\ref{fig:exp2_offloading}\text{(a)}. The DWM-PPO method demonstrates the fastest and most stable convergence, reaching a high-reward plateau $2000$ steps before Pure DWM and MADDPG. This accelerated convergence stems from a fundamental improvement in exploration quality: the latent refinement mechanism guides agents to more coordinated actions, enriching the experience buffer with higher-quality samples and allowing the agent to discover a superior policy more efficiently.

As shown in Fig.~\ref{fig:exp2_offloading}\text{(b)}, DWM-PPO achieves the highest spectral efficiency of $0.62~\text{bps/Hz}$, a $34.7\%$ improvement over the Pure DWM baseline ($0.46~\text{bps/Hz}$). This gain confirms the coordination mechanism effectively mitigates inter-agent interference. Moreover, the DQN and MLP-Bandit gate ($0.46~\text{bps/Hz}$ and $0.43~\text{bps/Hz}$, respectively) show no significant improvement over the non-coordinating Pure DWM, indicating their simpler offloading policies failed to learn an effective strategy.

Then, we analyze system reliability via the constraint violation rate in Fig.~\ref{fig:exp2_offloading}\text{(c)}. Our DWM-PPO method achieves the lowest violation rate of only $0.44\%$. This is a substantial reduction of over $40\%$ from the PureDWM's rate of $0.74\%$. This result powerfully highlights the mechanism's effectiveness in enhancing the agents' ability to satisfy stringent QoS and EH constraints. While a strong baseline like MAPPO also achieves a respectable violation rate of $0.9\%$, our method surpasses it in both reliability and spectral efficiency. The reason for this superior reliability lies in the synergy between the world model's predictive planning and the adaptive coordination, which allows agents to foresee and collectively navigate away from states that would lead to constraint breaches. In conclusion, the proposed offloading and refinement mechanism proves to be highly effective, delivering simultaneous improvements in learning speed, spectral efficiency, and, most importantly, system reliability.

\subsection{Scalability Analysis}
Finally, we evaluate the scalability and robustness of DWM-RO as the network density and interference complexity increase. We vary the number of FUEs (i.e., $K$) from 2 to 10, while keeping the number of SUEs fixed at 3. We compare the performance of our full DWM-PPO method against the strong MAPPO baseline and use a Random policy as a lower bound.

\begin{figure}[hbt!]
\vspace{-0.4cm}

    \centering

    \begin{subfigure}{0.9\columnwidth}
        \centering
        \includegraphics[width=\linewidth]{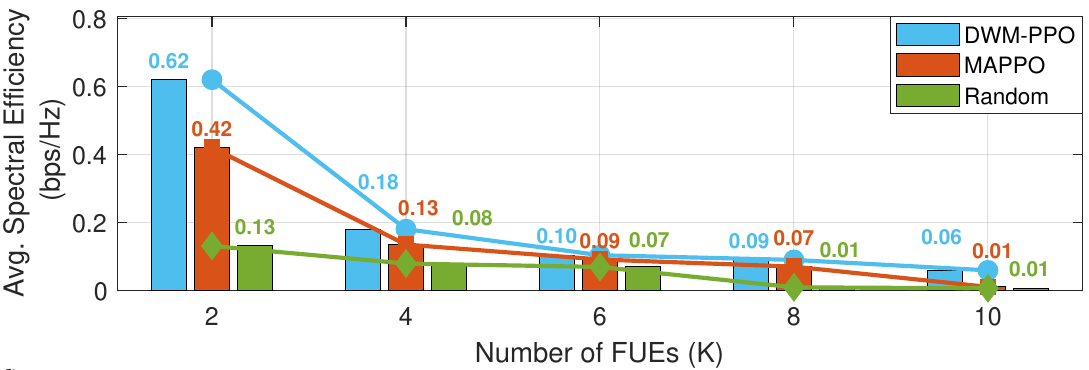} 
        \caption{\footnotesize Spectral Efficiency vs. Number of FUEs}
    \end{subfigure}
    
    \vspace{0cm} 

    \begin{subfigure}{0.9\columnwidth}
        \centering
        \includegraphics[width=\linewidth]{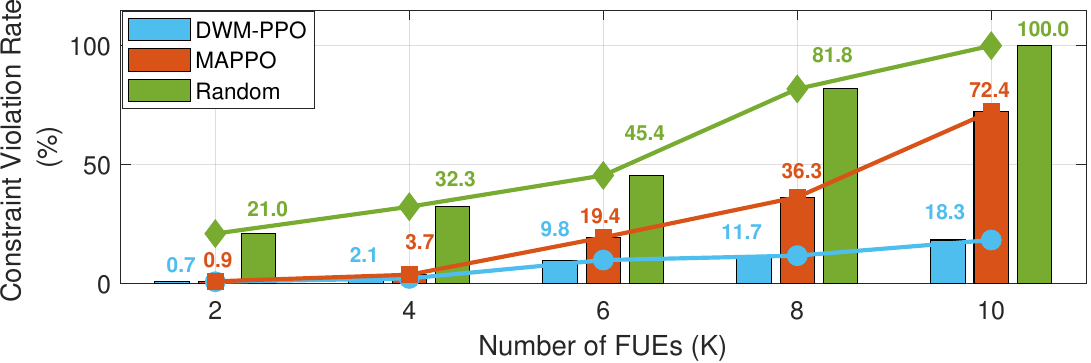} 
        \caption{ \footnotesize Constraint Violation Rate vs. Number of FUEs}
    \end{subfigure}

    \vspace{0cm} 

    \begin{subfigure}{0.9\columnwidth}
        \centering
        \includegraphics[width=\linewidth]{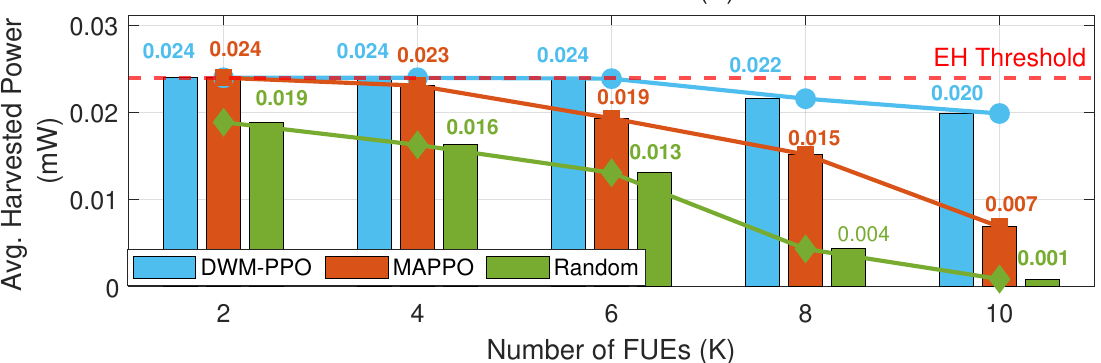} 
        \caption{\footnotesize Harvested Power vs. Number of FUEs}
    \end{subfigure}
    \caption{\footnotesize The scalability examination.}
    \label{fig:scalability}
\vspace{-0.3cm}
\end{figure}

{
The performance of all algorithms degrades as the network becomes more congested. As shown in Fig.~\ref{fig:scalability}, the average sum-rate for all methods decreases with more users due to the escalating co-channel interference. However, our DWM-PPO method consistently maintains the highest spectral efficiency across all values of $K$. Its performance degrades more gracefully, whereas MAPPO's throughput drops sharply when $K$ increases. This advantage comes at a modest cost in model size. Each DWM-RO agent requires approximately 0.85M parameters (3.4~MB), compared with 0.18M (0.7~MB) for a MAPPO agent. The additional parameters from the RSSM and gate networks enable the predictive planning and adaptive coordination that underpin the convergence speedup.

As shown in Fig.~\ref{fig:scalability}\text{(b)}, while DWM-PPO's constraint violation rate increases moderately with network density, it remains at a low level, reaching just under 20\% at $K=10$. In stark contrast, the violation rates for MAPPO and the Random policy escalate dramatically. At $K=10$, MAPPO's violation rate soars to over 70\%, and the Random policy fails completely with a 100\% violation rate. This result powerfully demonstrates the superior scalability of our approach.

The improved robustness lies in the adaptive nature of our framework. As $K$ increases, the interference environment becomes more chaotic, leading to higher uncertainty in each agent's local world model. This condition increasingly triggers the uncertainty-aware offloading gates. Recall that the gate reward in~\eqref{eq:gate_reward} includes a communication cost penalty $c$ that discourages unnecessary offloading. When $K$ is small (e.g., $K = 2$), co-channel interference remains moderate and agents can often resolve resource conflicts through local decision-making alone, so the coordination benefit frequently does not outweigh the cost $c$, resulting in a selective offloading rate of approximately 33\%. As $K$ grows, however, the escalating interference makes the coordination benefit far exceed the communication cost, and the gate learns to offload almost always. For $K = 10$, the offloading rate approaches 100\%. This adaptive behavior confirms that the gate has learned a meaningful policy that automatically balances coordination quality against communication overhead based on the prevailing network conditions. Consequently, the edge-side latent refinement mechanism plays an increasingly critical role in dense scenarios, actively managing the interference and coordinating agents' actions to maintain system feasibility.

Importantly, this increased coordination incurs negligible overhead at the FBS. The refinement mechanism introduces zero trainable parameters and requires only $\mathcal{O}(|\mathcal{O}(t)| \cdot d_z)$ floating-point operations. Even when all $K = 10$ agents offload simultaneously, this computation amounts to only $10 \times 32 = 320$ operations, completing in microseconds on any edge processor. The communication payload is equally lightweight. Each offloading event transmits only $\mathbf{z}_k(t) \in \mathbb{R}^{32}$ (128 bytes) on the uplink and receives $\tilde{\mathbf{z}}_k(t)$ on the downlink, yielding a round-trip of 256 bytes per agent. For $K = 10$, the worst-case aggregate payload is 2,560 bytes per time step, well within typical femtocell backhaul capacity. In contrast, standard MARL methods such as MAPPO, despite centralized training, struggle to generalize to the exponentially more complex interference patterns in dense scenarios, leading to a breakdown in reliability.

Finally, Fig.~\ref{fig:scalability}(c) shows that the average harvested power also decreases for all intelligent methods as $K$ increases. This is an expected outcome, as agents learn more conservative power control policies to mitigate the severe interference, thus reducing the total RF energy available in the environment. Even under these conditions, DWM-PPO demonstrates a much more efficient resource allocation strategy. At $K=10$, DWM-PPO maintains a high average harvested power of nearly $0.02~{\rm mW}$, whereas MAPPO's harvested power drops much more sharply to only $0.0069~{\rm mW}$. Moreover, DWM-PPO's superior ability to manage interference through adaptive coordination means its agents are not forced to reduce their transmit power as drastically as the MAPPO agents.
}

\subsection{Sensitivity to Imagination Horizon}

{
To examine the sensitivity of the proposed framework to the imagination rollout horizon $H$, we evaluate the prediction accuracy of the trained RSSM under open-loop rollouts of varying lengths. The choice of $H$ is important because it directly determines the planning depth of imagination-based policy optimization. A small $H$ limits the temporal horizon over which the actor-critic can anticipate future channel and interference evolution, while a large $H$ increases the accumulation of model bias and may inject inaccurate imagined returns into training. Starting from a real latent state, the model generates imagined trajectories of length $H$ using its learned dynamics and the same action sequence as the real environment, without access to new observations. The predicted observations and rewards are then compared against the ground truth.

\begin{figure}[t]
    \centering
    \includegraphics[width=0.9\columnwidth]{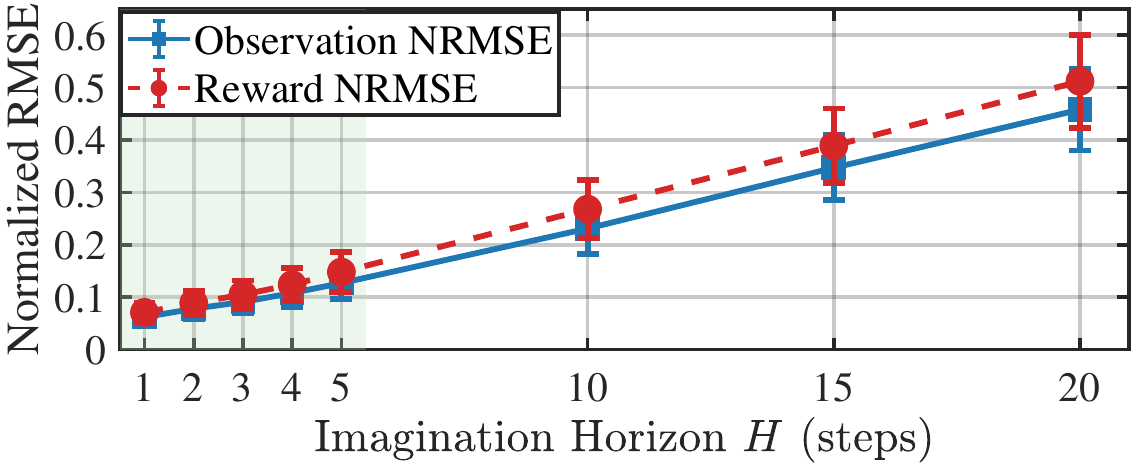}
    \caption{\footnotesize Normalized Root Mean Squared Error (NRMSE) of the RSSM's observation and reward predictions as a function of the imagination horizon $H$. The shaded region indicates the selected horizon $H=5$.}
    \label{fig:horizon_sensitivity}
\end{figure}

As shown in Fig.~\ref{fig:horizon_sensitivity}, both the observation and reward prediction errors remain low for $H \leq 5$, with the observation NRMSE staying below 0.13 and the reward NRMSE below 0.15. These values indicate that the imagined trajectories remain sufficiently close to the real dynamics to provide reliable multi-step supervision for actor-critic training. Beyond this range, the prediction error increases steadily due to the accumulation of model error over successive imagination steps. At $H = 20$, the observation NRMSE reaches 0.46, which corresponds to a large normalized deviation and indicates that the imagined trajectories have drifted substantially from reality. Therefore, although larger $H$ offers longer planning depth, the resulting model bias becomes too significant to support stable policy improvement. These results justify the choice of $H = 5$ as a balanced operating point between planning horizon and imagination fidelity.}

\section{Conclusion}
{
In this paper, we have addressed the complex problem of joint 
beamforming and power splitting in a SWIPT-enabled HetNet by 
proposing a novel hierarchical MARL framework, DWM-RO. Our 
framework integrates decentralized world models for proactive 
planning with an uncertainty-aware offloading gate and a 
lightweight latent refinement mechanism for efficient, adaptive 
coordination. Extensive experiments have demonstrated that 
DWM-RO significantly outperforms strong model-free baselines 
in terms of sample efficiency, spectral efficiency, and, most 
critically, constraint satisfaction. Furthermore, our framework 
exhibits superior scalability, maintaining robust performance 
in dense networks where conventional methods fail.

We acknowledge several limitations that point to promising 
directions for future work. First, the current single-FBS 
topology does not capture inter-cell interference in multi-cell 
networks. Extending DWM-RO to multi-FBS scenarios requires 
addressing hierarchical coordination across edge servers, 
where independently trained world models may produce latent 
representations that are not directly comparable, making 
cross-FBS latent aggregation a non-trivial open problem. 
Second, our framework assumes accurate local CSI. Incorporating 
robustness to CSI estimation errors through techniques such as 
domain randomization~\cite{liu2023wisr} or uncertainty-aware 
training~\cite{wang2025uncertainty} deserves further investigation. 
Third, although the per-agent computational cost is modest 
(0.85M parameters, 3.4~MB), deploying RSSM-based world models 
on resource-constrained FUE hardware remains a practical 
challenge that motivates future work on model compression 
techniques such as knowledge distillation~\cite{gou2021knowledge} and weight quantization.
}
\bibliographystyle{IEEEtran}
\bibliography{Ref}

@article{ha2018world,
  title={World models},
  author={Ha, David and Schmidhuber, J{\"u}rgen},
  journal={arXiv preprint arXiv:1803.10122},
  year={2018}
}

@article{hafner2019dream,
  title={Dream to control: Learning behaviors by latent imagination},
  author={Hafner, Danijar and Lillicrap, Timothy and others },
  journal={arXiv preprint arXiv:1912.01603},
  year={2019}
}

@article{hafnerdiverse,
  title={Mastering diverse domains through world models},
  author={Hafner, Danijar and Pasukonis, Jurgis and and others},
  journal={arXiv preprint arXiv:2301.04104},
  year={2023}
}

@article{hafner2025nature,
  title={Mastering diverse control tasks through world models},
  author={Hafner, Danijar and others},
  journal={Nature},
  pages={1--7},
  year={2025},
  publisher={Nature Publishing Group UK London}
}

@article{nomura2025decentralized,
  title={Decentralized collective world model for emergent communication and coordination},
  author={Nomura, Kentaro and others},
  journal={arXiv preprint arXiv:2504.03353},
  year={2025}
}

@article{lohani2016downlink,
  title={On downlink resource allocation for SWIPT in small cells in a two-tier HetNet},
  author={Lohani, Sudha and Hossain, Ekram and Bhargava, Vijay K},
  journal={IEEE Transactions on Wireless Communications},
  volume={15},
  number={11},
  pages={7709--7724},
  year={2016},
  publisher={IEEE}
}

@article{lu2018coordinated,
  title={Coordinated beamforming with artificial noise for secure SWIPT under non-linear EH model: Centralized and distributed designs},
  author={Lu, Yang and others},
  journal={IEEE Journal on Selected Areas in Communications},
  volume={36},
  number={7},
  pages={1544--1563},
  year={2018},
  publisher={IEEE}
}

@article{zhang2021joint,
  title={Joint coordinated beamforming and power splitting ratio optimization in {MU-MISO SWIPT}-enabled HetNets: A multi-agent {DDQN}-based approach},
  author={Zhang, Ruichen and others},
  journal={IEEE Journal on Selected Areas in Communications},
  volume={40},
  number={2},
  pages={677--693},
  year={2021},
  publisher={IEEE}
}

@ARTICLE{10322786,
  author={Lotfolahi, Amin and Ferng, Huei-Wen},
  journal={IEEE Transactions on Green Communications and Networking}, 
  title={A Multi-Agent Proximal Policy Optimized Joint Mechanism in mmWave HetNets With CoMP Toward Energy Efficiency Maximization}, 
  year={2024},
  volume={8},
  number={1},
  pages={265-278},
  doi={10.1109/TGCN.2023.3334495}}

@ARTICLE{10938906,
  author={Chen, Geng and others},
  journal={IEEE Transactions on Network Science and Engineering}, 
  title={MADDPG-M\&L: UAV-Assisted Joint User Association and Slicing Resource Allocation in HetNets}, 
  year={2025},
  volume={12},
  number={4},
  pages={2878-2894},
  doi={10.1109/TNSE.2025.3554991}}

@ARTICLE{8792117,
  author={Nasir, Yasar Sinan and Guo, Dongning},
  journal={IEEE Journal on Selected Areas in Communications}, 
  title={Multi-Agent Deep Reinforcement Learning for Dynamic Power Allocation in Wireless Networks}, 
  year={2019},
  volume={37},
  number={10},
  pages={2239-2250},
  doi={10.1109/JSAC.2019.2933973}}

@ARTICLE{8840901,
  author={Xu, Yongjun and others},
  journal={IEEE Internet of Things Journal}, 
  title={Robust Resource Allocation and Power Splitting in SWIPT Enabled Heterogeneous Networks: A Robust Minimax Approach}, 
  year={2019},
  volume={6},
  number={6},
  pages={10799-10811},
}

@ARTICLE{8891923,
  author={Zhang, Haijun and others},
  journal={IEEE Transactions on Wireless Communications}, 
  title={Energy Efficient Resource Management in SWIPT Enabled Heterogeneous Networks With NOMA}, 
  year={2020},
  volume={19},
  number={2},
  pages={835-845},
}

@article{worldmodel1,
  title={World Model-Based Learning for Long-Term Age of Information Minimization in Vehicular Networks},
  author={Lingyi Wang and others},
  journal={arXiv preprint arXiv:2505.01712},
  year={2025}
}

@article{worldmodel2,
  title={World Models for Cognitive Agents: Transforming Edge Intelligence in Future Networks},
  author={Changyuan Zhao and others},
  journal={arXiv preprint arXiv:2506.00417},
  year={2025}
}

@article{worldmodel3,
  title={MobiWorld: World Models for Mobile Wireless Network},
  author={Haoye Chai and Yuan Yuan and Yong Li},
  journal={arXiv preprint arXiv:2507.09462},
  year={2025}
}

@ARTICLE{9390169,
  author={Tataria, Harsh and others},
  journal={Proceedings of the IEEE}, 
  title={{6G} Wireless Systems: Vision, Requirements, Challenges, Insights, and Opportunities}, 
  year={2021},
  volume={109},
  number={7},
  pages={1166-1199},
  doi={10.1109/JPROC.2021.3061701}}

@ARTICLE{11195786,
  author={Hossain, Ekram and Vera-Rivera, Angelo},
  journal={IEEE Transactions on Technology and Society}, 
  title={{6G} Cellular Networks: Mapping the Landscape for the {IMT}-2030 Framework}, 
  year={2025},
  volume={},
  number={},
  pages={1-16},
  doi={10.1109/TTS.2025.3611364}}

@ARTICLE{9502719,
  author={Clerckx, Bruno and others},
  journal={IEEE Journal of Selected Topics in Signal Processing}, 
  title={Wireless Power Transfer for Future Networks: Signal Processing, Machine Learning, Computing, and Sensing}, 
  year={2021},
  volume={15},
  number={5},
  pages={1060-1094},
  doi={10.1109/JSTSP.2021.3098478}}

@article{pan2017performance,
  title={Performance analysis and optimization for SWIPT wireless sensor networks},
  author={Pan, Gaofeng and others},
  journal={IEEE Transactions on Communications},
  volume={65},
  number={5},
  pages={2291--2302},
  year={2017},
  publisher={IEEE}
}

@article{pan2016secrecy,
  title={On secrecy performance of MISO SWIPT systems with TAS and imperfect CSI},
  author={Pan, Gaofeng and others},
  journal={IEEE Transactions on Communications},
  volume={64},
  number={9},
  pages={3831--3843},
  year={2016},
  publisher={IEEE}
}

@ARTICLE{10032267,
  author={Zhang, Ruichen and others},
  journal={IEEE Journal on Selected Areas in Communications}, 
  title={Energy Efficiency Maximization in RIS-Assisted SWIPT Networks With RSMA: A PPO-Based Approach}, 
  year={2023},
  volume={41},
  number={5},
  pages={1413-1430},
  doi={10.1109/JSAC.2023.3240707}}

@article{clerckx2018fundamentals,
  title={Fundamentals of wireless information and power transfer: From RF energy harvester models to signal and system designs},
  author={Clerckx, Bruno and others},
  journal={IEEE Journal on Selected Areas in Communications},
  volume={37},
  number={1},
  pages={4--33},
  year={2018},
  publisher={IEEE}
}

@article{chu2025revolutionizing,
  title={Revolutionizing {6G}: Experimental Validation of an Optical Integrated Communication, Sensing, and Power Transfer System},
  author={Chu, Tiantian and others},
  journal={IEEE Journal on Selected Areas in Communications},
  year={2025},
  publisher={IEEE}
}

@ARTICLE{10354077,
  author={Zhang, Jiliang and others},
  journal={IEEE Transactions on Green Communications and Networking}, 
  title={Secrecy Analysis for NOMA-Based Multi-Antenna Satellite-UAV-Terrestrial SWIPT Systems}, 
  year={2024},
  volume={8},
  number={2},
  pages={672-685},
  doi={10.1109/TGCN.2023.3341874}}

@article{ng2014robust,
  title={Robust beamforming for secure communication in systems with wireless information and power transfer},
  author={Ng, Derrick and Lo, Ernest S and Schober, Robert},
  journal={IEEE Transactions on Wireless Communications},
  volume={13},
  number={8},
  pages={4599--4615},
  year={2014},
  publisher={IEEE}
}

@article{liu2024survey,
  title={A survey of recent advances in optimization methods for wireless communications},
  author={Liu, Ya-Feng and others},
  journal={IEEE Journal on Selected Areas in Communications},
  year={2024},
  publisher={IEEE}
}

@article{feriani2021single,
  title={Single and multi-agent deep reinforcement learning for AI-enabled wireless networks: A tutorial},
  author={Feriani, Amal and Hossain, Ekram},
  journal={IEEE Communications Surveys \& Tutorials},
  volume={23},
  number={2},
  pages={1226--1252},
  year={2021},
  publisher={IEEE}
}

@article{kang2023cooperative,
  title={Cooperative {UAV} resource allocation and task offloading in hierarchical aerial computing systems: A {MAPPO}-based approach},
  author={Kang, Hongyue and others},
  journal={IEEE Internet of Things Journal},
  volume={10},
  number={12},
  pages={10497--10509},
  year={2023},
  publisher={IEEE}
}

@inproceedings{liu2024efficient,
author = {Liu, Qihan and others},
title = {Efficient Multi-agent Reinforcement Learning by Planning},
year = {2024},
booktitle = {ICLR},
pages = {1–22},
}

@inproceedings{MABL,
author = {Venugopal, Aravind and others},
title = {{MABL}: Bi-Level Latent-Variable World Model for Sample-Efficient Multi-Agent Reinforcement Learning},
year = {2024},
booktitle = {AAMAS},
pages = {1865–1873},
}

@article{an2016secure,
  title={Secure transmission in cognitive satellite terrestrial networks},
  author={An, Kang and others},
  journal={IEEE Journal on Selected Areas in Communications},
  volume={34},
  number={11},
  pages={3025--3037},
  year={2016},
  publisher={IEEE}
}

@article{sharma2017performance,
  title={Performance analysis of overlay spectrum sharing in hybrid satellite-terrestrial systems with secondary network selection},
  author={Sharma, Pankaj K and others},
  journal={IEEE Transactions on Wireless Communications},
  volume={16},
  number={10},
  pages={6586--6601},
  year={2017},
  publisher={IEEE}
}

@article{zhou2023aerospace,
  title={Aerospace integrated networks innovation for empowering 6G: A survey and future challenges},
  author={Zhou, Di and others},
  journal={IEEE Communications Surveys \& Tutorials},
  volume={25},
  number={2},
  pages={975--1019},
  year={2023},
  publisher={IEEE}
}

@article{zhang2024generative,
  title={Generative AI agents with large language model for satellite networks via a mixture of experts transmission},
  author={Zhang, Ruichen and others},
  journal={IEEE Journal on Selected Areas in Communications},
  year={2024},
  publisher={IEEE}
}

@article{li2018robust,
  title={Robust chance-constrained secure transmission for cognitive satellite--terrestrial networks},
  author={Li, Bin and others},
  journal={IEEE Transactions on Vehicular Technology},
  volume={67},
  number={5},
  pages={4208--4219},
  year={2018},
  publisher={IEEE}
}

@article{zhang2013mimo,
  title={MIMO broadcasting for simultaneous wireless information and power transfer},
  author={Zhang, Rui and Ho, Chin Keong},
  journal={IEEE transactions on wireless communications},
  volume={12},
  number={5},
  pages={1989--2001},
  year={2013},
  publisher={IEEE}
}

@article{yue2023low,
  title={Low earth orbit satellite security and reliability: Issues, solutions, and the road ahead},
  author={Yue, Pingyue and others},
  journal={IEEE Communications Surveys \& Tutorials},
  volume={25},
  number={3},
  pages={1604--1652},
  year={2023},
  publisher={IEEE}
}

@ARTICLE{10949621,
  author={Liu, Yuan and others},
  journal={IEEE Transactions on Cognitive Communications and Networking}, 
  title={STAR-RIS Enabled ISAC Systems With RSMA: Joint Rate Splitting and Beamforming Optimization}, 
  year={2025},
  volume={},
  number={},
  pages={1-1},
  doi={10.1109/TCCN.2025.3558016}}

@article{xiao2002second,
  title={Second-order statistical properties of the WSS Jakes' fading channel simulator},
  author={Xiao, Chengshan and Zheng, Yahong R and Beaulieu, Norman C},
  journal={IEEE Transactions on communications},
  volume={50},
  number={6},
  pages={888--891},
  year={2002},
  publisher={IEEE}
}

@article{sherstinsky2020fundamentals,
  title={Fundamentals of recurrent neural network ({RNN}) and long short-term memory ({LSTM}) network},
  author={Sherstinsky, Alex},
  journal={Physica D: Nonlinear Phenomena},
  volume={404},
  pages={132306},
  year={2020},
  publisher={Elsevier}
}

@ARTICLE{8930071,
  author={Guo, Jing and others},
  journal={IEEE Transactions on Vehicular Technology}, 
  title={Performance of SWIPT for Full-Duplex Relay System With Co-Channel Interference}, 
  year={2020},
  volume={69},
  number={2},
  pages={2311-2315},
  doi={10.1109/TVT.2019.2958626}}

@inproceedings{fan2020theoretical,
  title={A theoretical analysis of deep Q-learning},
  author={Fan, Jianqing and others},
  booktitle={L4DC},
  pages={486--489},
  year={2020},
}

@inproceedings{haarnoja2018soft,
  title={Soft actor-critic: Off-policy maximum entropy deep reinforcement learning with a stochastic actor},
  author={Haarnoja, Tuomas and others},
  booktitle={ICML},
  pages={1861--1870},
  year={2018},
}

@article{gu2021proximal,
  title={Proximal policy optimization with policy feedback},
  author={Gu, Yang and others},
  journal={IEEE Transactions on Systems, Man, and Cybernetics: Systems},
  volume={52},
  number={7},
  pages={4600--4610},
  year={2021},
  publisher={IEEE}
}

@inproceedings{mo2025diffuse,
  title={Diffuse and Refine Latent Prior with Transformers For Neural ISP},
  author={Mo, Zhipeng and Li, Wenbo and Ding, Sihao},
  booktitle={ICIP},
  pages={1456--1461},
  year={2025},
}

@article{zhang2024decentralized,
  title={Decentralized transformers with centralized aggregation are sample-efficient multi-agent world models},
  author={Zhang, Yang and others},
  journal={arXiv preprint arXiv:2406.15836},
  year={2024}
}

@article{Kingma2014AdamAM,
  title={Adam: A Method for Stochastic Optimization},
  author={Diederik P. Kingma and Jimmy Ba},
  journal={CoRR},
  year={2014},
  volume={abs/1412.6980},
  url={https://api.semanticscholar.org/CorpusID:6628106}
}

@ARTICLE{10159432,
  author={Wang, Zining and others},
  journal={IEEE Communications Letters}, 
  title={Robust Beamforming for IRS-Aided SWIPT in Cognitive Satellite and Terrestrial Networks}, 
  year={2023},
  volume={27},
  number={9},
  pages={2408-2412},
  doi={10.1109/LCOMM.2023.3288483}}

@article{zhang2025comai,
  title={Com{AI}: The Convergence of Communication and Artificial Intelligence},
  author={Zhang, Ping and others},
  journal={IEEE Communications Surveys \& Tutorials},
  year={2025},
  publisher={IEEE}
}

@article{zhang2024spectrumnet,
  title={Generative ai on spectrumnet: An open benchmark of multiband 3d radio maps},
  author={Zhang, Shuhang and others},
  journal={IEEE Transactions on Cognitive Communications and Networking},
  year={2024},
  publisher={IEEE}
}

@article{liu2023wisr,
  title={WiSR: Wireless domain generalization based on style randomization},
  author={Liu, Shijia and others},
  journal={IEEE Transactions on Mobile Computing},
  volume={23},
  number={5},
  pages={4520--4532},
  year={2023},
  publisher={IEEE}
}

@article{wang2025uncertainty,
  title={Uncertainty awareness in wireless communications and sensing},
  author={Wang, Shixiong and others},
  journal={IEEE Communications Magazine},
  year={2025},
  publisher={IEEE}
}

@article{pan2022space,
  title={Space simultaneous information and power transfer: An enhanced technology for miniaturized satellite systems},
  author={Pan, Gaofeng and others},
  journal={IEEE Wireless Communications},
  volume={30},
  number={2},
  pages={122--129},
  year={2022},
  publisher={IEEE}
}

@article{wang2024safeguarding,
  title={Safeguarding inter-satellite transmissions: A viewpoint from covertness},
  author={Wang, Shuai and others},
  journal={IEEE Wireless Communications},
  volume={32},
  number={1},
  pages={221--228},
  year={2024},
  publisher={IEEE}
}

@article{ni2025llm,
  title={LLM aided spectrum-sharing LEO satellite communications},
  author={Ni, Zihan and others},
  journal={IEEE Journal on Selected Areas in Communications},
  year={2025},
  publisher={IEEE}
}

@article{gou2021knowledge,
  title={Knowledge distillation: A survey},
  author={Gou, Jianping and others},
  journal={International journal of computer vision},
  volume={129},
  number={6},
  pages={1789--1819},
  year={2021},
  publisher={Springer}
}
\end{document}